\documentclass[aps,prab,twocolumn,showpacs,preprintnumbers, superscriptaddress]{revtex4-2}

\usepackage{amsmath,amssymb,graphicx,siunitx}
\usepackage[colorlinks=true,linkcolor=blue,citecolor=blue,urlcolor=blue]{hyperref}
\usepackage{siunitx}
\usepackage{subcaption}
\usepackage{braket}
\usepackage{dashrule}
\usepackage{makecell}
\usepackage{setspace}
\usepackage{xcolor}
\usepackage{threeparttable}
\usepackage{multirow}

\newcommand{\dd}{\mathrm{d}}
\DeclareSIUnit{\nuclearmagneton}{\text{\ensuremath{\mu_N}}}

\begin{document}

\title{Eliminating beam-induced depolarizing effects in the hydrogen jet target \\ for high-precision proton beam polarimetry at the Electron-Ion Collider}

\author{F. Rathmann}
\thanks{Corresponding author. Email: frathmann@bnl.gov}
\affiliation{Brookhaven National Laboratory, Upton, NY 11973, USA}

\author{A. Nass}
\affiliation{Institut f\"ur Kernphysik, Forschungszentrum J\"ulich, 52425 J\"ulich, Germany}

\author{K.O. Eyser}
\affiliation{Brookhaven National Laboratory, Upton, NY 11973, USA}

\author{V. Shmakova}
\affiliation{Brookhaven National Laboratory, Upton, NY 11973, USA}

\author{E.C. Aschenauer}
\affiliation{Brookhaven National Laboratory, Upton, NY 11973, USA}

\author{G. Atoian}
\affiliation{Brookhaven National Laboratory, Upton, NY 11973, USA}

\author{A.\,Cannavo}
\affiliation{Brookhaven National Laboratory, Upton, NY 11973, USA}

\author{K. Hock}
\affiliation{Brookhaven National Laboratory, Upton, NY 11973, USA}

\author{H. Huang}
\affiliation{Brookhaven National Laboratory, Upton, NY 11973, USA}

\author{H.\,Lovelace}
\affiliation{Brookhaven National Laboratory, Upton, NY 11973, USA}

\author{G. Mahler}
\affiliation{Brookhaven National Laboratory, Upton, NY 11973, USA}

\author{J. Ritter}
\affiliation{Brookhaven National Laboratory, Upton, NY 11973, USA}

\author{G. Robert-Demolaize}
\affiliation{Brookhaven National Laboratory, Upton, NY 11973, USA}

\author{V. Schoefer}
\affiliation{Brookhaven National Laboratory, Upton, NY 11973, USA}

\author{P. Shanmuganathan}
\affiliation{Brookhaven National Laboratory, Upton, NY 11973, USA}

\author{E.\,Shulga}
\affiliation{Brookhaven National Laboratory, Upton, NY 11973, USA}

\author{H.\,Soltner}
\affiliation{Institute of Technology and Engineering, Forschungszentrum J\"ulich, 52425 J\"ulich, Germany}

\author{X.\,Chu}
\affiliation{Brookhaven National Laboratory, Upton, NY 11973, USA}

\author{Z.\,Zhang}
\affiliation{Brookhaven National Laboratory, Upton, NY 11973, USA}

\date{\today}

\begin{abstract}
We analyze beam-induced depolarizing effects in the hydrogen jet target (HJET) at the Relativistic Heavy Ion Collider (RHIC) that has been used for absolute hadron beam polarimetry and shall be employed at the Electron-Ion Collider (EIC). The EIC's higher bunch repetition frequencies and shorter bunch durations shift beam harmonics to frequencies that can resonantly drive hyperfine transitions in hydrogen, threatening to depolarize the target atoms.  Using frequency-domain analysis of beam harmonics and hyperfine transition frequencies, we establish a photon emission threshold above which beam-induced fields are too weak to cause significant depolarization. For EIC injection (\SI{23.5}{\GeV}) and flattop (\SI{275}{\GeV}), beam-induced depolarization through the bunch structure renders operation at the current RHIC magnetic guide field at the target of $B_0 = \SI{120}{\milli\tesla}$ untenable. Increasing the magnetic guide field at the target to $B_0 \approx \SI{400}{\milli\tesla}$ moves all hyperfine transition frequencies to at least three times the cutoff frequency, ensuring reliable absolute beam polarimetry with the required 1\% precision at the EIC.
\end{abstract}
		

\maketitle

\tableofcontents
	
\section{Introduction}
\label{sec:introduction}

The Electron-Ion Collider (EIC) is the next-generation facility designed to explore the internal structure of nucleons and nuclei with unprecedented precision\,\cite{osti_1765663}. By colliding polarized electrons with polarized protons and ions across a wide range of species and energies, the EIC will provide essential insights into the spin structure of the nucleon, the origin of mass, and the role of gluons in quantum chromodynamics\,\cite{Accardi2016, ABDULKHALEK2022122447}.

Accurate and reliable beam polarization measurements are essential to the success of the EIC scientific program. The polarized hadron running modes foresee operation with proton\,\cite{Zelenski:2018uzg} and helium-3 ($^3$He$^{++}$) beams\,\cite{ZELENSKI2023168494} and polarized electrons\,\cite{PhysRevAccelBeams.25.033401, Wang:2024bmt}, with the potential future addition of deuterons and other light ion species. A key performance requirement is to deliver beam polarization $P \geq 0.7$ with a relative uncertainty of $\left(\frac{\delta P}{P}\right)  \leq 1\%$\,\cite{ABDULKHALEK2022122447}. 

To meet these challenging requirements, the beam polarimetry shall characterize the full polarization vector $\vec{P} = (P_x, P_y, P_z)$, track the spatial profile of the polarization in the transverse planes\,\cite{PhysRevSTAB.15.041001} on a bunch-by-bunch basis, and monitor the polarization lifetime\,\cite{PhysRevAccelBeams.22.091001} throughout each store. For the EIC physics analyses described in Ref.\,\cite{ABDULKHALEK2022122447}, however, it is the projection of $\vec{P}$ onto the stable spin axis that matters, with any transverse (in-plane) polarization ideally minimized.

\begin{figure}[t]
	\centering
	\includegraphics[width = \columnwidth]{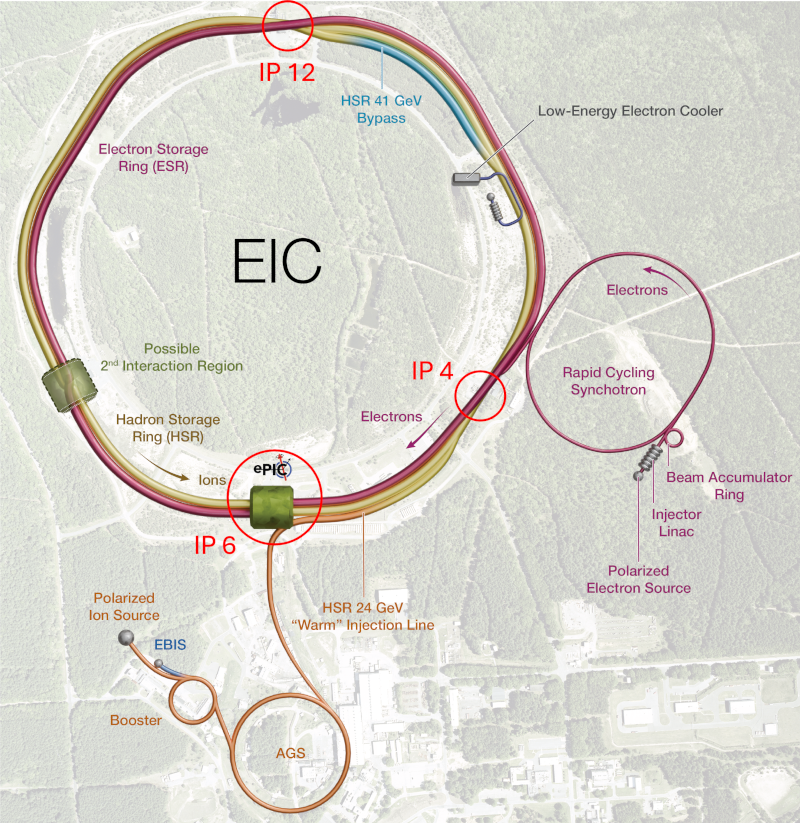}
	\caption{\label{fig:EIC-sketch}
		Aerial view of the Electron-Ion Collider (EIC) layout at Brookhaven National Laboratory. The primary detector, ePIC, is located at interaction point IP\,6 (6 o’clock position). For the EIC, the absolute HJET  polarimeter and one fast proton-carbon (pC) polarimeter will be installed at IP\,4 (4 o’clock), while an additional pC polarimeter is foreseen near IP\,6. During RHIC operation, the HJET and two pC polarimeters (one for each beam) were located at IP\,12 (12 o’clock). (Figure reflects the project planning status as of May 2025.)}
\end{figure}

The EIC polarimetry system will combine a high-accuracy absolute beam polarimeter, based on a polarized atomic beam and Breit-Rabi polarimeter (BRP), with fast relative proton-carbon (pC) polarimeters for bunch-by-bunch monitoring of polarization profiles and  beam lifetime. The polarized jet target and two pC polarimeters\,\cite{Huang:2006cs} for horizontal and vertical measurements are pres\-ent\-ly installed at RHIC’s interaction point (IP) 12, where they have been successfully operated throughout the spin program\,\cite{ ALEKSEEV2003392,RHICPolarimetryGroup2018}. For the EIC, these instruments will be relocated to IP\,4 (4 o’clock position), while a second pC polarimeter will be deployed at IP\,6\,\cite{ABDULKHALEK2022122447}, collocated with the primary detector (ePIC) and between the spin rotators, as illustrated in Fig.\,\ref{fig:EIC-sketch}.

It should be noted that the EIC polarimetry requirements represent a substantial enhancement over current RHIC capabilities, as the polarized hydrogen jet target (HJET) was designed to achieve an absolute calibration of the proton-carbon polarimeters to approximately 5\%\,\cite{ALEKSEEV2003392}. The stringent 1\% relative polarization uncertainty requirement demanded by the EIC physics program necessitates a comprehensive reassessment of all systematic effects, including the beam-induced target depolarizing mechanisms analyzed in this work. 

Beam-induced depolarizing effects due to the bunch structure of the beam, as observed at the HERMES polarized storage cell target in the HERA ring\,\cite{PhysRevLett.82.1164}, pose a significant risk to polarized target operations at the EIC. This paper quantitatively assesses such effects under the anticipated EIC beam and optics conditions at IP4, with the goal of ensuring reliable operation of the polarized target and enabling absolute beam polarimetry. The comparison to RHIC operation at IP12 serves as a benchmark to identify and understand depolarizing mechanisms that may arise at the EIC. The EIC is expected to operate with substantially enhanced beam conditions at both injection and flattop energies, particularly in bunch number (10 $\times$
higher), bunch length (10 $\times$ shorter), and stored beam current (3 $\times$ higher), necessitating separate analyses for EIC injection and flattop conditions.

The paper is organized as follows. Section\,\ref{sec:abs-beam-polarimetry} outlines the principle of absolute beam polarimetry using the HJET and the CNI scattering method. Section\,\ref{sec:hfs-of-hydrogen} reviews the hyperfine level structure of hydrogen, the transition frequencies, and the target operation at RHIC. Section\,\ref{sec:fourier-analysis} analyzes the temporal and spectral properties of beam-induced magnetic fields. Section\,\ref{sec:rhic-depol} provides a detailed analysis of beam-induced depolarization effects at RHIC flattop, including resonance conditions, photon emission thresholds, and spatial magnetic field distributions at the target. Section\,\ref{sec:EIC-depol} extends this analysis to the EIC at both injection and flattop, examining how higher bunch frequencies and different beam parameters affect depolarization of hydrogen atoms when operated at the same holding field as at RHIC, and presents a solution for reliable EIC operation. Section\,\ref{sec:conclusion} offers concluding remarks.

\section{Principle of absolute beam polarimetry}
\label{sec:abs-beam-polarimetry}

\subsection{Analyzing power in the CNI region}

At the beam energies available at the Alternating Gradient Synchrotron (AGS) and RHIC, no scattering processes exist for which the analyzing power $A_y$ is known with sufficient precision to achieve the beam polarization uncertainty of $\left( \frac{\delta P}{P} \right) \le 1\%$\,\cite{Haeberli:2005tj, Poblaguev:2019saw}. The method developed at RHIC for absolute beam polarization measurements therefore relies on a polarized atomic beam source (ABS) combined with a BRP\,\cite{Roser:AnnRev:2002, Haeberli:2005tj}. This technique enables an accurate determination of the target polarization $Q$, which is then used to calibrate the beam polarization based on measured asymmetries in elastic proton-proton scattering in the Coulomb-nuclear interference (CNI) region\,\cite{PhysRevD.48.3026, PhysRevD.79.094014,Buttimore2013}.

The CNI asymmetry arises from the interference between electromagnetic and hadronic amplitudes at small momentum transfer\,\cite{ALEKSEEV2003392, Kopeliovich:1974ee, PhysRevD.18.694, PhysRevD.59.114010}. This same electromagnetic amplitude also governs the proton’s magnetic moment $\mu_p = g_p \mu_N = 2(1 + G_p)\, \mu_N$, where $g_p$ is the proton magnetic g-factor, $G_p = (g_p - 2)/2$ is the anomalous gyromagnetic ratio\footnote{This can be rewritten as $\mu_p = 2(1 + G)\,\mu_N = 2\,\mu_N + 2G\,\mu_N$, showing explicitly that the anomalous contribution $\mu_p^\text{anom.} = 2G\,\mu_N$ reflects the deviation from the Dirac value $2\,\mu_N$.}. The nuclear magneton $\mu_N = e\hbar / 2m_p$ and related constants are listed in Table~\ref{tab:Bc-constants}. 

At high energies, such as those at RHIC, the CNI region provides a maximum analyzing power of $A_y \approx 0.046$ at $t = \SI{0.003}{GeV\squared}$ for $pp$ elastic scattering\,\cite{PhysRevD.48.3026, Poblaguev:2019saw}. The role of electromagnetic interference in determining $A_y$ and enabling absolute polarization calibration has been emphasized, e.g., in Ref.\,\cite{Buttimore2013}. Because the absolute magnitude of $A_y$ depends on both theoretical modeling and experimental normalization, an accurately calibrated polarized target (via ABS and BRP) remains essential for achieving high-precision absolute beam polarization determination at the EIC.
	
\begin{table*}[htb]
	\centering
	\caption{\label{tab:Bc-constants} Fundamental physical constants and hydrogen-specific parameters used for analyzing hyperfine structure and beam-induced depolarization effects.}
	\begin{ruledtabular}
		\begin{tabular*}{\textwidth}{@{\extracolsep{\fill}}llllr}
			Quantity & Symbol & Value & Unit & Reference \\
			\hline
			Hyperfine frequency of hydrogen & $f_\mathrm{hfs}$ & \num{1.420405748e9} & \si{\hertz} & \cite{Diermaier:2016fsy} \\
			Boltzmann constant & $k_B$ & \num{1.380649e-23} & \si{\joule\per\kelvin} & \cite{CODATA2018} \\
			Hydrogen atom mass & $m_\text{H}$ & \num{1.6735575e-27} & \si{\kilogram} & \cite{CODATA2018} \\
			Gyromagnetic ratio of H (electron) & $\gamma_\text{H}/2\pi$ & \num{28.025e9} & \si{\hertz\per\tesla} & \cite{CODATA2018} \\
			Planck constant & $h$ & \num{6.62607015e-34} & \si{\joule\second} & \cite{CODATA2018} \\
			Elementary charge & $e$ & \num{1.602176634e-19} & \si{\coulomb} & \cite{CODATA2018} \\
			Permeability of free space & $\mu_0$ & $4\pi \times 10^{-7}$ & \si{\henry\per\meter} & \cite{CODATA2018} \\
			Electron mass & $m_e$ & \num{9.1093837015e-31} & \si{\kilogram} & \cite{CODATA2018} \\
			Proton mass & $m_p$ & \num{1.67262192369e-27} & \si{\kilogram} & \cite{CODATA2018} \\
			Bohr magneton & $\mu_B = \frac{e\hbar}{2m_e}$ & \num{5.7883818e-5} & \si{\electronvolt\per\tesla} & \cite{CODATA2018} \\
			Nuclear magneton & $\mu_N = \frac{e\hbar}{2m_p}$ & \num{3.1524513e-8} & \si{\electronvolt\per\tesla} & \cite{CODATA2018} \\
			Electron $g$-factor & $g_J$ & \num{2.0023193} & -- & \cite{CODATA2018} \\
			Proton $g$-factor & $g_I$ & \num{5.5856947} & -- & \cite{CODATA2018} \\
		\end{tabular*}
	\end{ruledtabular}
\end{table*}

\subsection{Polarized hydrogen target setup at IP12 in RHIC}

\begin{figure}[htb]
	\centering
	\includegraphics[width=0.8\columnwidth]{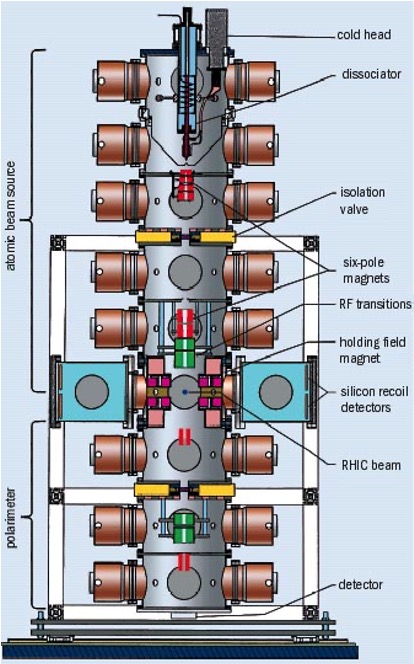}
	\caption{	\label{fig:HJET} Schematic layout of the HJET polarimeter, taken from Ref.\,\cite{Haeberli:2005tj}, showing the atomic beam source, the scattering chamber, and the Breit–Rabi polarimeter. The detector geometry and coordinate system are detailed in Fig.\,\ref{fig:hjet-schematic}.}
\end{figure}

The HJET polarimeter\,\cite{Haeberli:2005tj,Zelenski2005}, presently located at IP12 in RHIC (see Fig.\,\ref{fig:EIC-sketch}), consists of three core components that operate together as an integrated system. These include the polarized ABS, a scattering chamber with a holding field magnet, and the BRP, all arranged along a common vertical axis as illustrated in Fig.\,\ref{fig:HJET}. Recoil protons are detected in the horizontal plane, perpendicular to the directions of the circulating beams. 

The  system operates under a shared vacuum maintained by nine identical cylindrical chambers, each measuring \SI{50}{\centi\meter} in diameter and \SI{32}{\centi\meter} in length. The dissociator chamber is evacuated by three turbomolecular pumps, each of the subsequent chambers is evacuated by a pair of turbomolecular pumps in a nine-stage differential pumping system, with each individual pump providing a pumping speed of \SI{1000}{\ell\per\second} and a compression ratio of \num{e6} for H$_2$.

The ABS generates a polarized hydrogen atomic beam with a target thickness of approximately \SI{1e12}{atoms\per\centi\meter\squared}\,\cite{Zelenski2005}, enabling continuous, non-invasive operation without disturbing the circulating beams or generating background for other experiments. While the initial design aimed to achieve a beam polarization uncertainty of $\left( \frac{\delta P}{P} \right)\le 5\%$\,\cite{bunce-annrev:2000}, recent work reported in Ref.\,\cite{POBLAGUEV2020164261} claimed substantial reductions in systematic uncertainties to $\left(\frac{\delta P}{P}\right)_\text{syst} \le 0.5\%$. However, the methodology applied in Ref.\,\cite{POBLAGUEV2020164261} for determining the molecular content of the atomic beam is inappropriate and underestimates the contribution of hydrogen molecules in the target. Data from the ANKE ABS at COSY\,\cite{MIKIRTYCHYANTS201383}, analyzed in Appendix\,\ref{app:mol-to-atoms}, show that the molecular content in an atomic beam is on the order of 3 to 4\%, consistent with findings in\,\cite{NASS2003633, Nass:2002xj}, and contradicting the claims made in Ref.\,\cite{POBLAGUEV2020164261}.

The present study evaluates the modifications necessary for adapting the HJET polarimeter system to the EIC environment, where significantly higher beam currents and increased bunch repetition frequencies present new challenges compared to RHIC, with the goal of achieving a relative systematic uncertainty of $\left( \frac{\delta P}{P}\right) \leq 1\%$. While additional modifications may be required, the adaptations identified in this study are definitively necessary for successful operation under EIC conditions.

\subsection{Absolute polarization calibration}

\begin{figure}[htb]
	\centering
	\includegraphics[width=0.95\columnwidth]{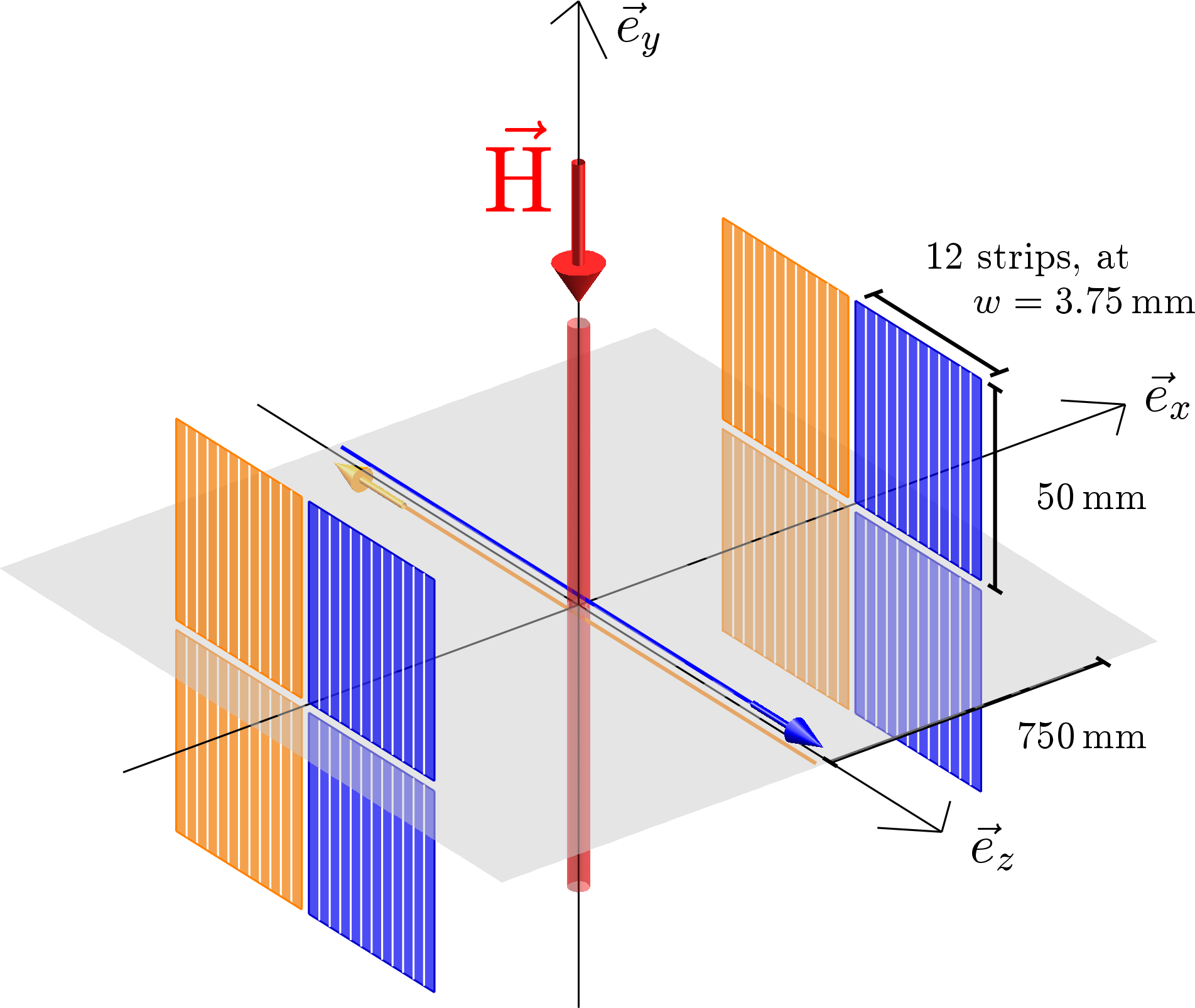}
	\caption{	\label{fig:hjet-schematic} Sketch of the detector setup at the HJET at RHIC. The atomic $\vec{\text{H}}$ beam enters from above and intersects the hadron beams orthogonally. Recoil protons are detected using silicon strip detectors placed symmetrically to the left and right of the vertically separated blue and yellow beams. 8 Si strip detectors are used with 12 vertical strips, each with a pitch of $w = \SI{3.75}{mm}$, and \SI{500}{\micro \meter} thickness. The coordinate system is indicated with $\vec e_x \parallel$ to ring plane, $\vec e_y \perp$ to ring plane, and $\vec e_z$ along the beam momentum.}

\end{figure}

The polarized atomic beam intersects the circulating hadron beam in a vacuum chamber equipped with silicon strip detectors positioned on both sides of the beam axis, as illustrated in Fig.\,\ref{fig:hjet-schematic}. The blue detector pair measures the scattering asymmetry of the blue beam, and the yellow pair does the same for the yellow beam. From these scattering asymmetries, the vertical beam polarization component $P_y$ is extracted\,\cite{POBLAGUEV2020164261}.

With the present setup of detectors to the left (L) and right (R) of the beams at IP\,12 in RHIC (Fig.\,\ref{fig:hjet-schematic}), and a magnetic guide field of 
\begin{equation}
	\vec B_0 = B_0\cdot \vec e_y\,,
	\label{eq:B0}
\end{equation}
where $ B_0 \approx \SI{120}{\milli \tesla}$, the vertical beam polarization component $P_y$ can be absolutely determined in the CNI region near $\theta_\text{cm} = \SI{90}{\degree}$ based on the target polarization $Q_y$, determined by the  BRP. The relation governing the beam polarization dependence of scattered protons is given by\footnote{Here the Madison convention is used\,\cite{Ohlsen1972, MadisonConvention1971}, which employs Cartesian observables ($A_y$, $A_{yy}$, $A_{xx}$, etc.) relative to the beam and scattering plane, rather than the spherical tensor notation ($A_N$, $A_{NN}$, etc.) of the Ann Arbor convention\,\cite{Ashkin:1977ek}. The Madison convention is more natural for fixed target experiments with inherently Cartesian coordinate systems.}
\begin{equation}
	\sigma(\theta, \phi) = \sigma_0(\theta) \left[1 + A_y(\theta) P_y \cos \phi \right]\,,
\end{equation}
where $\theta$ denotes the scattering angle, $\sigma_0$ is the unpolarized cross section, $\phi$ is the azimuthal scattering angle, and $A_y$ is the corresponding analyzing power.  When the sign of the vertical target polarization $Q_y$ is periodically reversed to compensate for asymmetries caused by differences in the detector geometry or detector efficiency in the L and R directions\,\cite{Hanna1965}, the target asymmetry is determined from the accumulated number of counts in the detectors via  
\begin{equation}
	\epsilon_\text{target} = \frac{\text{L} - \text{R}}{\text{L} + \text{R}} = A_y \, Q_y\,.
\end{equation}
A measurement of the corresponding asymmetry with beam particles determines $\epsilon_\text{beam}$. In elastic $pp$ scattering, and more general in the elastic scattering of identical particles, $A_y$ is the same regardless of which particle is polarized. The beam polarization $P_y$ is then obtained from
\begin{equation}
	P_y = \frac{\varepsilon_\text{beam}}{\varepsilon_\text{target}} \cdot Q_y\,.
\end{equation}

When beam and target particles are both polarized, detector systems with full azimuthal coverage provide access to the other two components of the beam polarization $P_x$ and $P_z$, as established in, e.g., \,\cite{PhysRevC.58.658, PhysRevC.58.1897}. Obviously, with an unpolarized target, due to parity conservation as in, e.g., proton-proton scattering, the longitudinal beam polarization component $P_z$ cannot be directly measured.

The polarimeters envisioned for proton beams at the EIC will combine a high-precision absolute polarimeter, based on an ABS and a BRP, with two fast and flexible relative pC polarimeters in IP4 and IP6. While the polarized hydrogen jet target technology developed for RHIC provides a proven foundation, the substantially higher beam intensities and bunch repetition frequencies at the EIC necessitate a comprehensive reassessment of beam-induced depolarization effects and a refined experimental design. This includes both the achievement of a beam polarization measurement to a precision of $\left( \frac{\Delta P}{P}\right) \leq 1\%$ and the capability to determine the complete beam spin vector $\vec P $. Other critical aspects, such as the determination of the absolute nuclear target polarization using the BRP with the accuracy required for achieving the above beam polarization precision, will be addressed in forthcoming work.

\section{The hyperfine structure of hydrogen}
\label{sec:hfs-of-hydrogen}

The hydrogen atom's hyperfine structure arises from the magnetic interaction between the proton and electron spins. This coupling creates an energy landscape that is exquisitely sensitive to external magnetic fields -- both static and time-varying. Understanding this structure is essential because beam-induced RF fields can resonantly drive transitions between these levels, potentially destroying the nuclear polarization that the target provides for absolute beam polarimetry. The beam bunch structure generates time-varying electromagnetic fields that can resonantly drive hyperfine transitions in the hydrogen target, leading to depolarization of the target atoms.

\subsection{Breit-Rabi energy levels and field dependence}
\label{sec:BR-energy-levels}

In the absence of an external magnetic field, the ground state of hydrogen exhibits hyperfine structure due to the interaction between the electron and nuclear spins\, \cite{ramsey1956, Haeberli1967, CohenTannoudjiQM-German}, resulting in two energy levels: a higher-energy triplet state with total angular momentum $F = 1$ (threefold degenerate with $m_F = -1, 0, +1$) and a lower-energy singlet state with $F = 0$ ($m_F = 0$). When an external magnetic field is applied, the degeneracy of the $F = 1$ level is lifted through the Zeeman effect, splitting it into three distinct energy levels corresponding to the three possible values of $m_F$. The $F = 0$ state, having no magnetic moment in the coupled representation, shifts in energy but remains a single level. This magnetic field-induced splitting transforms the original two-level system into the four energy levels $|1\rangle$, $|2\rangle$, $|3\rangle$, and $|4\rangle$.

These four hyperfine states can be precisely defined in the uncoupled basis $\{|m_J, m_I\rangle\}$ where both the electron and nuclear spin projections $m_J, m_I = \pm \frac{1}{2}$ are specified independently,
\begin{equation}
	\begin{split}
		\left|1\right\rangle &= \left|+\frac{1}{2}, +\frac{1}{2}\right\rangle = |e^\uparrow p^\uparrow\rangle \quad (m_F = +1) \\
		\left|2\right\rangle &= \left|+\frac{1}{2}, -\frac{1}{2}\right\rangle = |e^\uparrow p^\downarrow\rangle \quad (m_F = 0) \\
		\left|3\right\rangle &= \left|-\frac{1}{2}, -\frac{1}{2}\right\rangle = |e^\downarrow p^\downarrow\rangle \quad (m_F = -1) \\
		\left|4\right\rangle &= \left|-\frac{1}{2}, +\frac{1}{2}\right\rangle = |e^\downarrow p^\uparrow\rangle \quad (m_F = 0)\,,
	\end{split}
	\label{eq:HFS-state-desription-1-to-4}
\end{equation}
where $m_F = m_J + m_I$ is the total magnetic quantum number, and the arrow notation indicates the relative orientation of electron ($e$) and nuclear ($p$) spins. States $|1\rangle$ and $|3\rangle$ have definite total angular momentum $F = 1$ with $m_F = +1$ and $m_F = -1$, respectively, while states $|2\rangle$ and $|4\rangle$, both having $m_F = 0$, form a coupled system that mixes under the influence of external magnetic fields.

The energy levels of these states in an external magnetic field can be quantitatively described by the Breit-Rabi formula \cite{PhysRev.38.2082.2}. For an atom with total electron angular momentum $J = \frac{1}{2}$ and nuclear spin $I = \frac{1}{2}$, the energy levels are given by
\begin{equation}
	E_{F, m_F}(B) = -\frac{E_\mathrm{hfs}}{4} + g_I \mu_N m_I B 
	\pm \frac{E_\mathrm{hfs}}{2} \sqrt{1 + 2 m_F x + x^2}\,,
	\label{eq:Breit-Rabi-formula}
\end{equation}
where $E_\mathrm{hfs} = h \cdot f_\mathrm{hfs}$ is the zero-field hyperfine splitting, $g_I$ is the nuclear g-factor of the proton, $\mu_N$ is the nuclear magneton, and $m_I = \pm \frac{1}{2}$ is the nuclear spin projection and $m_F$ is the magnetic quantum number of the total angular momentum $F$. The $\pm$ sign corresponds to the $F = 1$ (upper sign) and $F = 0$ (lower sign) hyperfine levels. The dimensionless field strength parameter $x$ is defined as
\begin{equation}
	x = \frac{g_J \mu_B B}{E_\mathrm{hfs}}\,,
	\label{eq:def-of-x}
\end{equation}
where $g_J$ is the electron g-factor and $\mu_B$ is the Bohr magneton (see Table\,\ref{tab:Bc-constants} for numerical values). The first term in Eq.\,\eqref{eq:Breit-Rabi-formula} represents the zero-field energy offset, the second term describes the nuclear Zeeman effect (interaction of the nuclear magnetic moment with the external field), and the square root term captures the combined hyperfine and electron Zeeman interactions.

The ground-state hyperfine splitting in hydrogen is known with exceptional precision. A recent measurement yielded
\begin{equation}
	f_\mathrm{hfs} = (1420405748.4 \pm 3.4_\mathrm{stat} \pm 1.6_\mathrm{syst})\,\si{Hz}\,,
\end{equation}
as reported in Ref.\,\cite{Diermaier:2016fsy}.  In energy units, using the measured hyperfine frequency $f_\mathrm{hfs}$ and Planck’s constant $h$ from Table~\ref{tab:Bc-constants}, the hyperfine splitting energy is given by
\begin{equation}
	E_\text{hfs} = \frac{h f_\mathrm{hfs}}{e} = \SI{5.87432617e-6}{\electronvolt}
	\label{eq:EHFS}
\end{equation}

The magnetic field $B_\text{c}$ at which the Zeeman interaction equals the hyperfine interaction (i.e., $x = 1$) is
\begin{equation}
	B_c = \frac{E_\mathrm{hfs}}{g_J \mu_B} 
	\approx \SI{50.684}{\milli \tesla}\,,
	\label{eq:Bc}
\end{equation}
where the CODATA 2018\,\cite{CODATA2018} values from Table\,\ref{tab:Bc-constants} for $h$, $e$, and $m_e$ were used and the classical definition $\mu_B = e\hbar / (2 m_e)$.

\begin{figure}[t]
	\centering
	\includegraphics[width= \columnwidth]{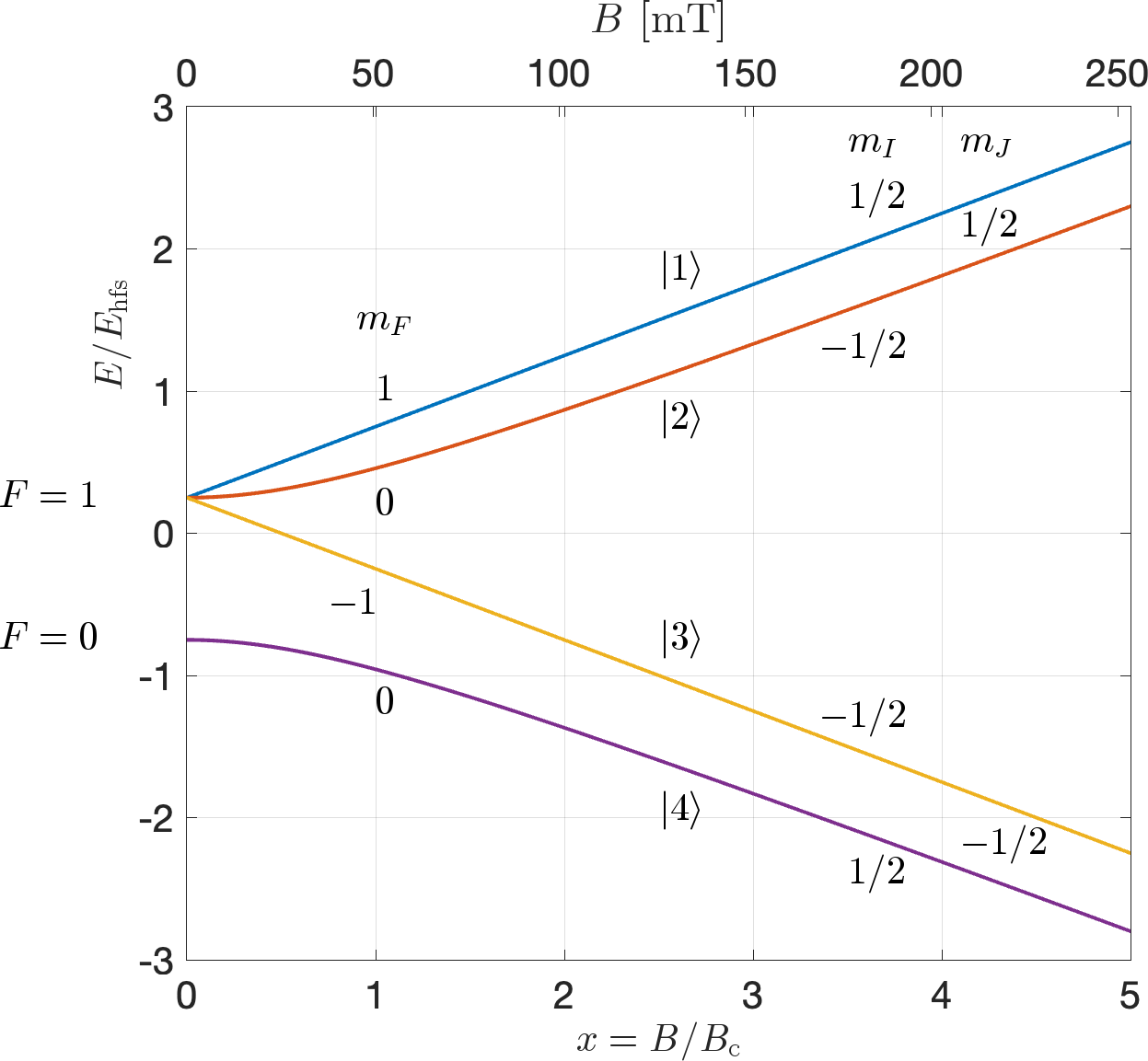}
	\caption{	\label{fig:HFS-H} Hyperfine energy levels of hydrogen $|1\rangle$ to $|4\rangle$ labeled with their quantum numbers $F$, $m_F$, $m_I$, $m_J$  vs. magnetic field, using Eq.~(\ref{eq:hyperfine-energies}) with $E_\text{hfs}$ from Eq.~(\ref{eq:EHFS}) and $B_c$ from Eq.~(\ref{eq:Bc}). The bottom axis is in units of $x = B/B_c$, the top axis gives $B$ in mT.
	}

\end{figure}

For the simplified energy expressions that follow, the nuclear Zeeman term $g_I \mu_N m_I B$ in Eq.\,\eqref{eq:Breit-Rabi-formula} is omitted since it is negligible compared to the hyperfine and electron Zeeman interactions (the nuclear magneton is approximately 1836 times smaller than the Bohr magneton). The hyperfine energies, whose complete derivation is presented in \ref{app:Breit-Rabi-derivation}, can be written as
\begin{equation}
	\begin{split}
		E_{|1\rangle}(x) & = \frac{E_\mathrm{hfs}}{2} \left( -\frac{1}{2} + (1 + x) \right) \,, \\
		E_{|2\rangle}(x) & = \frac{E_\mathrm{hfs}}{2} \left( -\frac{1}{2} + \sqrt{1 + x^2} \right), \\
		E_{|3\rangle}(x) & = \frac{E_\mathrm{hfs}}{2} \left( -\frac{1}{2} + (1 - x) \right) \,, \\
		E_{|4\rangle}(x) & = \frac{E_\mathrm{hfs}}{2} \left( -\frac{1}{2} - \sqrt{1 + x^2} \right)\,,
	\end{split}
	\label{eq:hyperfine-energies}
\end{equation}
where the different states are labeled according to their total and magnetic quantum numbers $|F, m_F\rangle$, as shown in Fig.\,\ref{fig:HFS-H}. As the external field increases, the relevant quantum numbers change from the coupled representation $F, m_F$ to the uncoupled basis $m_I, m_J$.  The expressions in Eq.\,\eqref{eq:hyperfine-energies} are valid for all magnetic field strengths, transitioning smoothly from the weak-field Zeeman regime ($x \ll 1$) through the intermediate regime to the strong-field Paschen-Back limit ($x\gg 1$). In the high-field (Paschen--Back) limit, the eigenstates effectively become pure product states of nuclear and electron spin projections.

\begin{figure}[htb]
	\centering
	\includegraphics[width= \columnwidth]{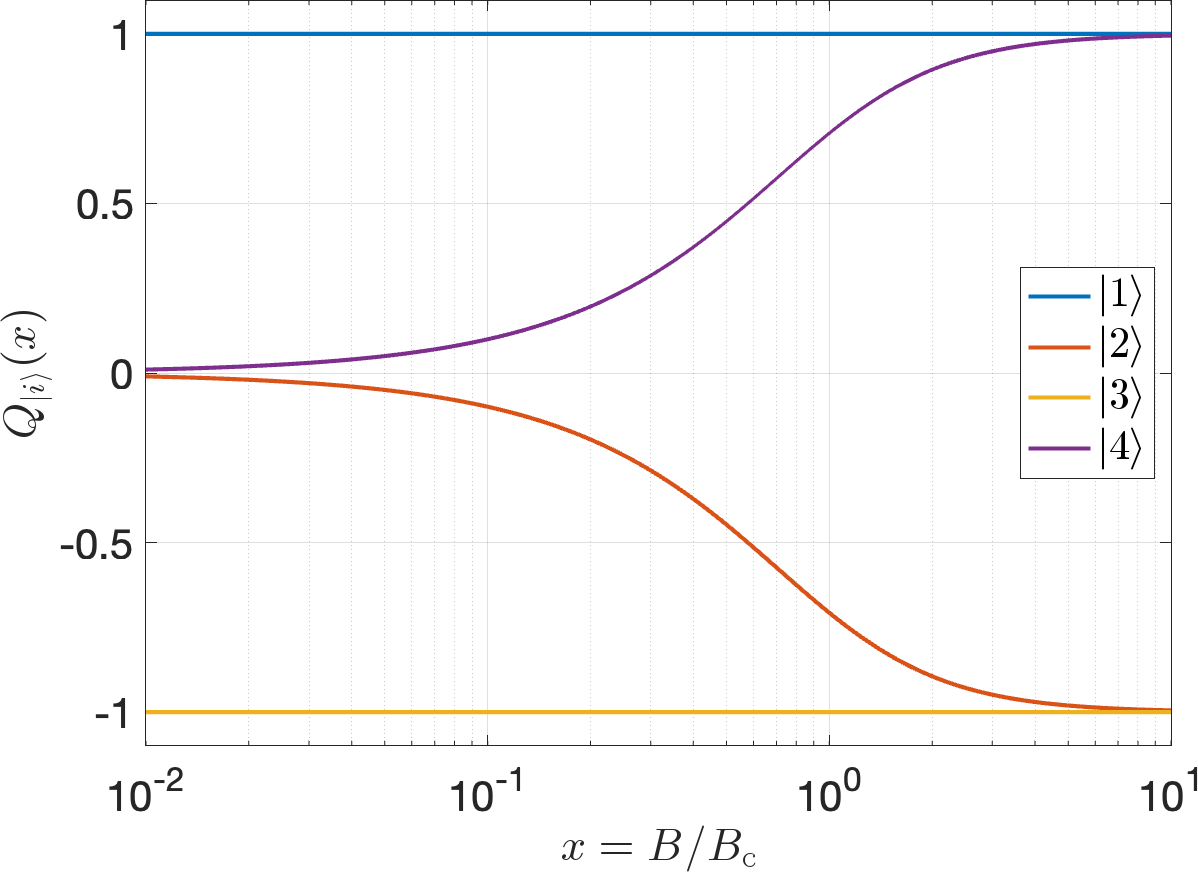}
   \caption{	\label{fig:nuclear-polarization} Nuclear target polarization of hydrogen hyperfine states as a function of the dimensionless magnetic field parameter $x = B/B_c$, as given by Eqs.~\eqref{eq:Q-of-x-for1to4}. States $|1\rangle$ and $|3\rangle$ maintain constant nuclear polarizations of $+1$ and $-1$, respectively, at all field strengths. The mixed states $|2\rangle$ and $|4\rangle$ exhibit field-dependent polarizations that evolve from zero in the weak-field limit ($x \to 0$) to $\pm 1$ in the strong-field limit ($x \to \infty$).}
\end{figure}

The nuclear target polarization of each hyperfine state also depends on the magnetic field strength through the parameter $x$. As derived in \ref{app:Breit-Rabi-derivation}, the field-dependent nuclear polarizations are given by 
\begin{equation}
	\begin{split}
		Q_{|1\rangle}(x) &= +1 \quad \text{(constant)}\,, \\
		Q_{|2\rangle}(x) &= -\frac{x}{\sqrt{1 + x^2}}\,, \\
		Q_{|3\rangle}(x) &= -1 \quad \text{(constant)}\,, \\
		Q_{|4\rangle}(x) &= +\frac{x}{\sqrt{1 + x^2}}\,,
	\end{split}
	\label{eq:Q-of-x-for1to4}
\end{equation}
and are depicted in Fig.\,\ref{fig:nuclear-polarization}. 
States $|1\rangle$ and $|3\rangle$ maintain constant nuclear polarizations of $+1$ and $-1$, respectively, while the mixed states $|2\rangle$ and $|4\rangle$ exhibit field-dependent polarizations that evolve from zero in the weak-field limit to $\pm 1$ in the strong-field limit. 

%

\subsection{Hyperfine transition frequencies in hydrogen}
\label{subsec:transition-frequencies}

As the magnetic field increases, the energies of the hyperfine states evolve, leading to field-dependent transition frequencies between them. The energies $E_{|i\rangle}(B)$ entering these transitions are given by the parametrization in Eq.~(\ref{eq:hyperfine-energies}), expressed as a function of the dimensionless parameter $x$, defined in Eq.\,\eqref{eq:def-of-x}. The transition frequency between two hyperfine states $|i\rangle$ and $|j\rangle$ is then given by
\begin{equation}
	f_{ij}(B) = \frac{E_{|i\rangle}(B) - E_{|j\rangle}(B)}{h}\,.
	\label{eq:transition-frequency}
\end{equation}

There are six allowed transitions between the four hyperfine states. Following the classification scheme\footnote{A similar classification scheme was used in the analysis of beam-induced depolarization at the HERMES experiment at DESY, where RF fields generated by the HERA electron beam lead to transitions between hyperfine states of hydrogen and deuterium atoms in the polarized storage cell gas target\,\cite{PhysRevLett.82.1164,Airapetian2005}.} introduced by Ramsey\,\cite[p.\,242]{ramsey1956}, they are grouped according to the orientation of the RF field $B_1$ relative to the static magnetic field $B_0$ and the associated selection rules:
\begin{itemize}
	\item \textbf{$\pi$-transitions} ($B_1 \perp B_0$): These occur within the same $F$ multiplet and obey $\Delta F = 0$, $\Delta m_F = \pm 1$. The two $\pi$-transitions are:
	\begin{itemize}
		\item $f_{12}^{\pi}$: between $|1\rangle$ and $|2\rangle$
		\item $f_{23}^{\pi}$: between $|2\rangle$ and $|3\rangle$
	\end{itemize}
	
	\item \textbf{$\sigma$-transitions} ($B_1 \parallel B_0$): These occur between different $F$ multiplets and satisfy $\Delta F = \pm 1$, $\Delta m_F = 0, \pm 1$. The three $\sigma$-transitions are:
	\begin{itemize}
		\item $f_{14}^{\sigma}$: between $|1\rangle$ and $|4\rangle$
		\item $f_{24}^{\sigma}$: between $|2\rangle$ and $|4\rangle$
		\item $f_{34}^{\sigma}$: between $|3\rangle$ and $|4\rangle$
	\end{itemize}
	
	\item \textbf{Two-photon transition} ($\Delta m_F = 2$): Forbidden as a single-photon process due to selection rules, this transition can occur through two-photon absorption:
	\begin{itemize}
		\item $f_{13}^{2\gamma}$: between $|1\rangle$ and $|3\rangle$
	\end{itemize}
\end{itemize}

\begin{figure}[t]
	\centering
	\includegraphics[ width= \columnwidth]{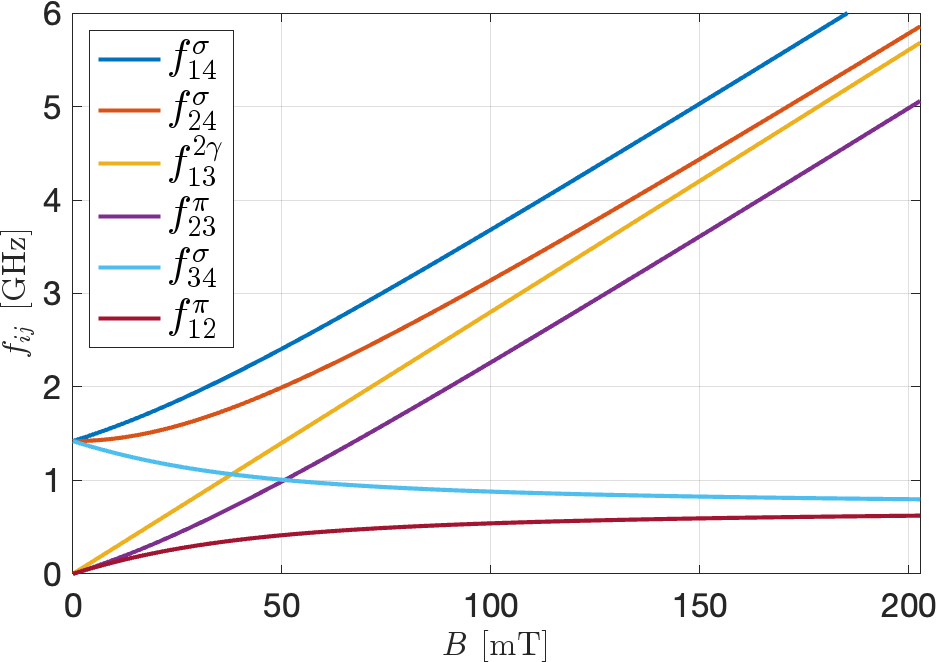}
	\caption{	\label{fig:transitions} Magnetic-field dependence of the transition frequencies $f_{ij}(B)$ between the hydrogen hyperfine states, calculated using Eq.\,\eqref{eq:transition-frequency}. The transitions are labeled as $f_{ij}^{\pi}$, $f_{ij}^{\sigma}$, or $f_{ij}^{2\gamma}$ according to their selection rules and field orientation. All frequencies are shown in GHz as a function of the magnetic field up to $4B_c$.}

\end{figure}

These six transition frequencies, representing all possible transitions between the four hyperfine states, are plotted in Fig.\,\ref{fig:transitions} as a function of magnetic field up to $4B_c$. 

\subsection{RHIC hydrogen jet target operation}

In the polarized hydrogen jet target, atoms are prepared in specific hyperfine state combinations by the atomic beam source, typically $|1\rangle + |4\rangle$ or $|2\rangle + |3\rangle$. These particular combinations are chosen because they maximize atomic beam intensity while maintaining high polarization, as the nuclear  polarization components of these states are nearly identical, allowing efficient population of both states without significant polarization loss.

The RHIC hydrogen jet target operates at a nominal holding field of $B_0 = 120$~mT ($\approx 2.4B_c$), placing it in the regime where hyperfine and Zeeman interactions are comparable. The  efficiencies (or transmissions) of the atomic hyperfine states being  transported in the magnetic focusing system of the source to the interaction point depend on the effective magnetic moments\footnote{The effective magnetic moment can be defined as $\mu_\text{eff} = -\frac{dE}{dB}$, where $E$ is the energy of a state from the Breit-Rabi formula in Eq.\,\eqref{eq:Breit-Rabi-formula}.}. The BRP measures the relative populations of the hyperfine states in the beam to determine the target polarization. Thus states $|2\rangle$ and $|4\rangle$, which have field-dependent effective magnetic moments (as evident from the varying slopes in Fig.\,\ref{fig:HFS-H}), experience different transmission efficiencies in the ABS compared to states $|1\rangle$ and $|3\rangle$ with constant effective magnetic moments, altering the target polarization even under idealized conditions. Any process that redistributes these populations -- such as beam-induced RF transitions -- directly affects the nuclear target polarization and thus the accuracy of absolute proton beam polarimetry. The transition frequencies calculated above establish which RF field components from the circulating beam can resonantly drive such depolarizing transitions.

\section{Temporal evolution and spectral properties of beam-in\-duced magnetic fields at RHIC}
\label{sec:fourier-analysis} 

Electromagnetic fields generated by the circulating beam bunches are the primary drivers of potential depolarization in the hydrogen target, as they can resonantly excite hyperfine transitions when their frequency components match the transition frequencies discussed in Section 3.

For the RHIC analysis presented in this section, we focus exclusively on flattop operation at \SI{255}{GeV} for two practical reasons. First, there is very limited experimental data available for nuclear target polarization measurements at injection energy due to insufficient statistics, whereas at flattop the polarized hydrogen target has been operated continuously throughout the typically 8-hour store duration. Second, the transverse beam size at injection is generally larger than at flattop by approximately a factor of $ \approx \sqrt{\gamma_{\text{flat}}/\gamma_{\text{inj}}} \approx  \sqrt{255\,\text{GeV}/23.5\,\text{GeV}} \approx \sqrt{11}$, resulting in correspondingly smaller magnetic field amplitudes at the target location. The flattop analysis therefore represents the more critical scenario and establishes a well-characterized benchmark for comparison with the EIC conditions analyzed in  Section\,\ref{sec:EIC-depol}. 

The analysis proceeds by first characterizing the temporal structure of individual bunches and the resulting periodic pulse train, then deriving the frequency-domain spectrum that determines which hyperfine transitions can be resonantly driven by the beam-induced fields.

\subsection{Bunch time structure and pulse shape}

At RHIC, the circulating beam is composed of $N_\text{b} = 120$ equally spaced bunches, each containing approximately $N_p = 2 \times 10^{11}$ protons. For the present discussion, the abort gap is neglected. The longitudinal profile of each individual bunch is  approximated by a Gaussian current distribution in time,
\begin{equation}
	I_\text{b}(t) = \frac{Q_\text{b}}{\sqrt{2\pi} \sigma_t} \exp\left( -\frac{t^2}{2\sigma_t^2} \right),
	\label{eq:gaussian-bunch}
\end{equation}
where $Q_\text{b} = N_p e$ is the total bunch charge and $\sigma_t$ is the temporal width of the bunch. For RHIC at top energy, the bunch length is approximately $\sigma_L = \SI{0.55}{m}$ in the lab frame, which yields a time-domain width of
\begin{equation}
	\sigma_t = \frac{\sigma_L}{\beta c},
	\label{eq:sigma_t}
\end{equation}
with $\beta \approx 1$. This corresponds to a temporal bunch width of $\sigma_t \approx \SI{1.84}{ns}$, and, using Eq.\,\eqref{eq:gaussian-bunch}, a peak current of a single bunch of $I_\text{b}^{\text{pk}} = Q_\text{b} / (\sqrt{2\pi} \sigma_t) \approx \SI{6.97}{A}$ for  RHIC flattop parameters.

\begin{figure}[t]
	\centering
	\begin{subfigure}[b]{\columnwidth}
		\centering
		\includegraphics[ width=\textwidth]{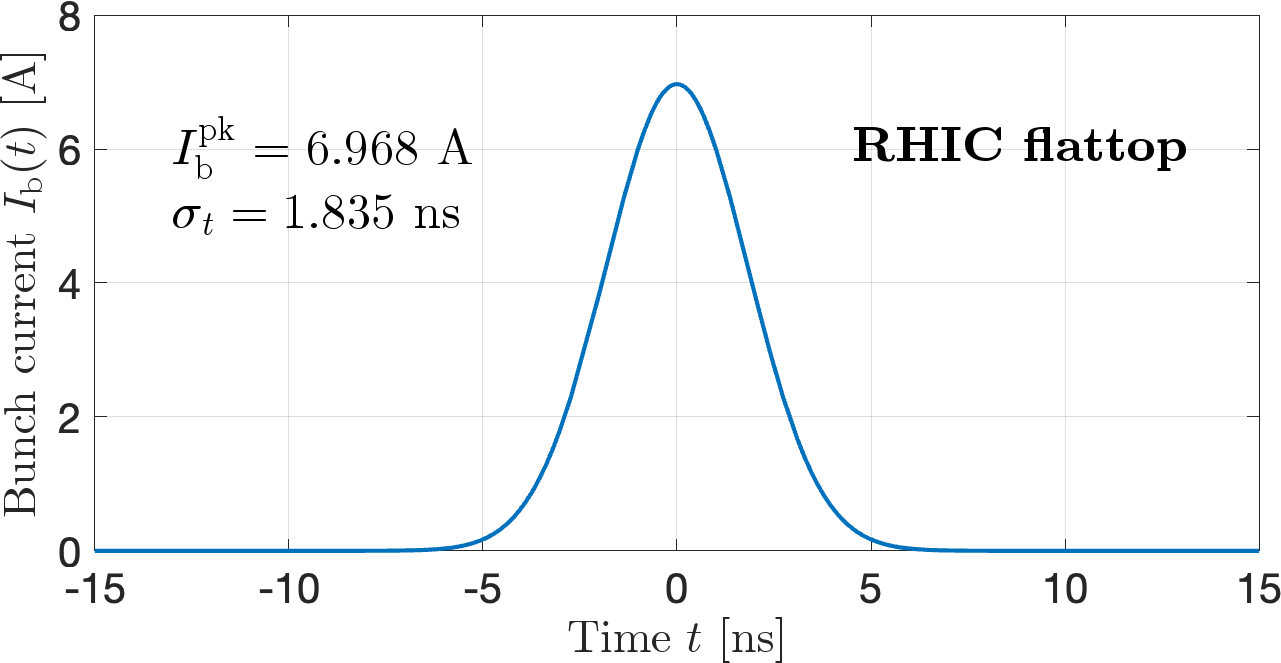}
		\caption{Single Gaussian bunch current profile $I_\text{b}(t)$  from Eq.\,(\ref{eq:gaussian-bunch}) with Gaussian width $\sigma_t$ and bunch charge $Q$, as listed in Table\,\ref{tab:bunch-params}.}
		\label{fig:gaussian-bunch}
	\end{subfigure}
	\hfill
	\begin{subfigure}[b]{\columnwidth}
		\centering
		\includegraphics[ width=\textwidth]{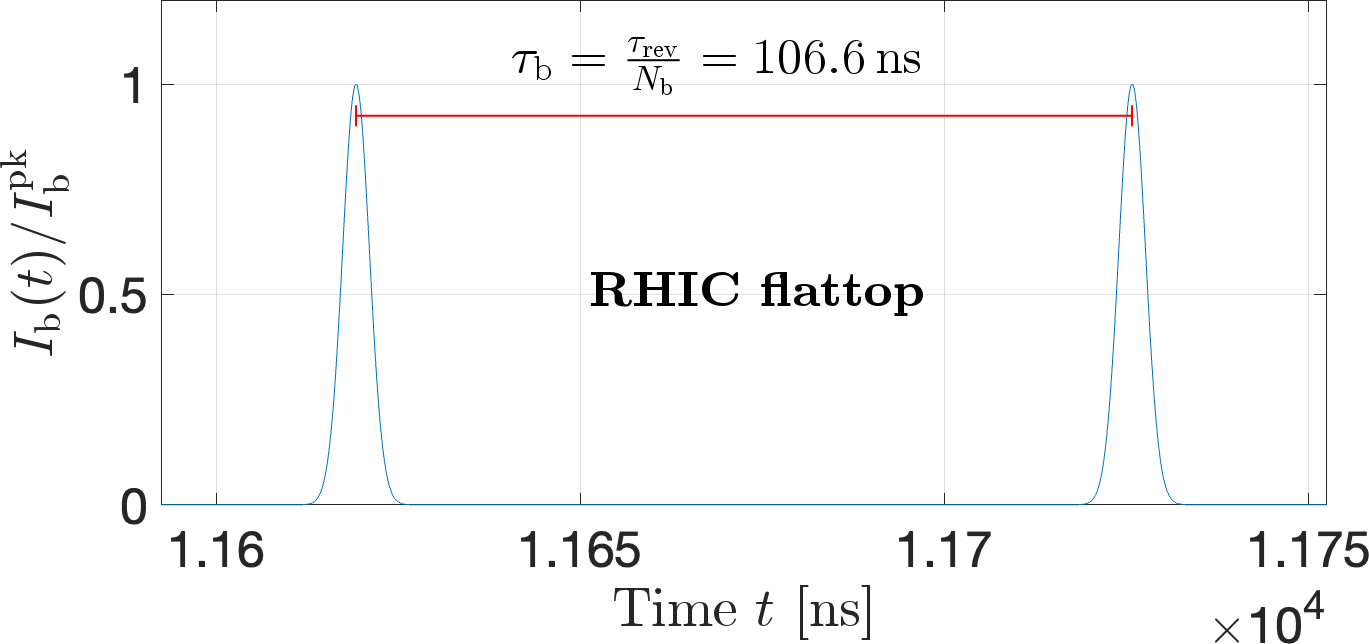}
		\caption{Two consecutive Gaussian bunches, separated by the nominal bunch spacing $\tau_\text{b}$.}
		\label{fig:gaussian-two-bunch}
	\end{subfigure}
	\caption{Temporal current profiles of RHIC bunches on flattop at \SI{255}{GeV}. 
	Panel (a): shape of an individual Gaussian bunch used in modeling the longitudinal current distribution. Panel (b): periodic repetition of the bunch shape with the nominal bunch spacing $\tau_\text{b}$. }
	\label{fig:bunch-profiles}
\end{figure}

The full set of machine and bunch parameters is summarized in Table~\ref{tab:bunch-params}. A graphical representation of the bunch current profile is shown in Fig.\,\ref{fig:gaussian-bunch}, illustrating the temporal shape used in subsequent frequency-domain analyses. Figure\,\ref{fig:gaussian-two-bunch} shows two consecutive RHIC  bunches at flattop and their temporal spacing.

\begin{table*}[htbp]
	\centering
	\singlespacing  
	\begin{threeparttable}		
		\caption{		\label{tab:bunch-params} Beam bunch and machine parameters for RHIC flattop and EIC injection and flattop nominal conditions. The average beam current $I_{\text{avg}}$ corresponds to the equivalent DC current that would deliver the same total charge flow as the bunched beam circulating at revolution frequency $f_{\text{rev}}$. The bottom part lists the transverse beam parameters at the HJET locations in IP\,12 (RHIC) and IP\,4 (EIC) that is used to evaluate the magnetic field $B(r)$ from the bunch current distribution.}
		\begin{ruledtabular}
		\begin{tabular}{llr|r|rr}
			\multicolumn{3}{c|}{}  & RHIC at IP\,12 & \multicolumn{2}{c}{EIC at IP\,4}\\
			
			Quantity & Symbol / Definition & Unit  & flattop & injection & flattop\\
			\hline
			
			Total beam energy                    & $E_{\text{beam}}$ & \si{GeV}         & 255               & $23.5$ & $275$ \\
			
			Lorentz factor (lab) & $\beta$ & -- &  $1.0000$ & $0.9992$ & $1.0000$\\
			Lorentz factor (lab) & $\gamma$ & -- & $\num{271.7762}$  & $25.0460$ & $293.0920$\\
			Protons per bunch                   & $N_p$             & \num{e10}           &  $20$ &  \num{27.6} & \num{6.9}  \\
			Bunch charge & $Q_\text{b} = N_p e$ & \si{nC} & \num{32.044}  & 44.220 & 11.055 \\
			
			Number of bunches                   & $N_\text{b}$             & --               & $120$                & $290$ & $1160$ \\
			
			Circumference                  & $L$ & m    &   \multicolumn{3}{c}{\makecell[c]{\hdashrule[0.75ex]{2.45cm}{0.5pt}{2pt 0pt} \num{3833.85} \hdashrule[0.75ex]{2.45cm}{0.5pt}{2pt 0pt}}}   \\
			
			Bunch length (RMS)                  & $\sigma_L$        & \si{m}       & 0.55              & $0.24$ & $0.06$ \\
			Temporal bunch width (RMS)         & $\sigma_t = \sigma_L / (\beta c)$ & \si{ns} & 1.835              & 0.801  & 0.200 \\
			
			Peak current (per bunch) & $I^\text{pk}_\text{b} = Q_\text{b} / (\sqrt{2\pi}\, \sigma_t)$ & \si{A} & 6.968 & 22.019 & 22.036 \\
			
			Revolution time & $\tau_{\text{rev}} = L / (\beta c)$ & $\si{\micro\second}$ &  $\num{12.792}$ & 12.802 &12.792  \\
			
			Revolution frequency & $f_{\text{rev}} = 1 / \tau_{\text{rev}}$ & \si{kHz} & 78.175 & 78.113 & 78.175  \\
			
			Bunch spacing                      & $\tau_\text{b} = \tau_{\text{rev}} / N_\text{b}$ & \si{ns} & 106.598  & 44.144 &   11.027 \\
			
			Bunch frequency                    & $f_\text{b} = 1 / \tau_\text{b}$ & \si{MHz}       & 9.381             & 22.653  & 90.683\\
			
			Average beam current               & $I_{\text{avg}} = N_\text{b} N_p e f_{\text{rev}}$ & \si{A} & 0.301 & 1.002 & 1.003  \\\hline
			Normalized rms emittance (horizontal)  & $\epsilon_x^\text{n}$ & \si{\micro \meter} & $2.5$ & $3.3$ & $3.3$ \\
			Normalized rms emittance (vertical)  & $\epsilon_y^\text{n}$ & \si{\micro \meter} & $2.5$  & $0.3$  & $0.3$\\
			Normalized average rms emittance   & $\epsilon_\text{avg}^\text{n} = \sqrt{\epsilon_x^\text{n} \cdot \epsilon_y^\text{n}}$ & \si{\micro \meter} & $2.5$  & $0.995$& $0.995$\\
			\hline
			Beta function (horizontal) & $\beta_x$ & m & $5.340$\tnote{a} & $93.600$\tnote{b} &$230.323$\tnote{b} \\
			Beta function (vertical) & $\beta_y$ & m& $6.190$\tnote{a} & $39.590$\tnote{b} & $69.935$\tnote{b}  \\
			Average beta function & $\beta_\text{avg} = \sqrt{\beta_x \beta_y} $ & m &  $5.749$ & $60.874$ & $126.916$\\\hline
			Transverse rms beam size (horizontal) & $\sigma_x = \sqrt{\beta_x \, \epsilon^\text{n}_x / (\beta \gamma)}$ & mm & -- & $3.513$ & 1.610\\
			Transverse rms beam size (vertical) & $\sigma_y = \sqrt{\beta_y \, \epsilon^\text{n}_y / (\beta \gamma)} $ & mm & -- & $0.689$ & 0.268\\
			Transverse 95\% beam size (horizontal) & $\sigma_x^{95}= \sigma_x \cdot \sqrt{5.993}$ & mm & -- & $8.600$& 3.942 \\		
			Transverse 95\% beam size (vertical) & $\sigma_y^{95} = \sigma_y \cdot \sqrt{5.993}$ & mm & -- & $1.686$ & 0.655 \\
			Radial rms beam size & $\sigma_r= \sqrt{\sigma_x \sigma_y}$ & mm & $0.23$ & $1.566$ & 0.656\\
			Radial beam size (95\%) & $\sigma_r^{95} = \sigma_r \cdot \sqrt{5.993}$ & mm & $0.56$ & $3.808$ & $1.607$\\
		\end{tabular}
		\end{ruledtabular}
		\begin{tablenotes}
			\footnotesize
			\item[a] In RHIC run 22, the $\beta$ functions at the location of the HJET in IP\,12 were determined by Guillaume Robert-Demolaize in January 2022\,\href{http://www.cadops2.bnl.gov/elogs/entryList.jsp?DATABY=day&ELOG=RHIC&DATE=01/18/2022&DIR=none#1565516}{(link)}.
			\item[b] Values for the future location of the HJET in IP\,4 were generated by Henry Lovelace\,III for flattop (July 2024) and injection (May 2025).
		\end{tablenotes}	
	\end{threeparttable}
\end{table*}

\subsection{Modeling the bunch train as a periodic source}

We begin by analyzing the frequency content of the bunch current and the resulting RF magnetic field spectrum. 

Each individual bunch is described by a temporal current distribution $I_\text{b}(t)$ as shown in Fig.\,\ref{fig:gaussian-bunch}. The full beam current $I(t)$ as seen by a stationary observer is modeled as a convolution of the single-bunch profile with a comb of delta functions spaced by the bunch interval $\tau_\text{b}$ via
\begin{equation}
	I(t) = I_\text{b}(t) \ast \sum_{n=-\infty}^{\infty} \delta(t - n\tau_\text{b})\,.
	\label{eq:bunch-train-convolution}
\end{equation}
The symbol $\ast$ denotes the convolution operator, defined for two functions $f(t)$ and $g(t)$ as
\begin{equation}
	(f \ast g)(t) = \int_{-\infty}^{\infty} f(t') \, g(t - t')\, \mathrm{d}t' \,,
	\label{eq:convolution-def}
\end{equation}
where $t'$ is a dummy integration variable. In the present context, this operation replicates the single-bunch current profile $I_\text{b}(t)$ at each multiple of the bunch spacing $\tau_b$, producing a periodic pulse train with a harmonic structure that reflects the bunch frequency $f_\text{b} = 1 / \tau_\text{b}$. Understanding this temporal structure is essential for analyzing the beam-induced radiofrequency fields that can depolarize the atoms in the target. 

\subsection{Frequency-domain spectrum of the beam}

We now determine the time structure of the circulating beam and its harmonic content by extending the single-bunch description to a periodic bunch train.

The total time-dependent current $I(t)$ of the circulating RHIC beam at flattop is constructed as a convolution of the single-bunch current profile $I_\text{b}(t)$ with a Dirac comb $\sum_n \delta(t - n\tau_\text{b})$ of period $\tau_\text{b}$, as given in Eq.~\eqref{eq:bunch-train-convolution}. The convolution of a localized function with a delta train yields a periodic pulse train of the same shape, replicated every $\tau_\text{b}$.

Due to the periodicity of the resulting current signal, the spectral content consists of harmonics of the bunch frequency $f_\text{b} = 1 / \tau_\text{b}$, modulated by the Fourier transform of the individual bunch shape.

\subsubsection*{Analytical form of the Gaussian bunch spectrum}

The Fourier transform of the Gaussian current distribution from Eq.~\eqref{eq:gaussian-bunch} is well known and yields a Gaussian in the frequency domain, given by
\begin{equation}
	\tilde{I}_\text{b}(f) = I_\text{b}^\text{pk} \cdot \exp\left(-2\pi^2 f^2 \sigma_t^2\right),
	\label{eq:gaussian-ft}
\end{equation}
where $f$ is the frequency and $\sigma_t$ the bunch width. This can also be written as $\tilde{I}_\text{b}(f) = I_\text{b}^\text{pk} \cdot \exp(-f^2/2\sigma_f^2)$ with the frequency-domain width $\sigma_f = 1/(2\pi\sigma_t)$. This expression describes the envelope of the spectral intensity of the bunch pulse train, falling off exponentially with frequency. The full spectrum of the periodic train is thus given by
\begin{equation}
	\tilde{I}(f) = \tilde{I}_\text{b}(f) \cdot \sum_{n=-\infty}^{\infty} \delta(f - n f_\text{b})\,.
\end{equation}

\subsubsection*{Numerical evaluation of the Fourier spectrum}

To compare this analytical result with a numerical calculation, the bunch train signal $I(t)$ was sampled over a time window of $2\tau_{\text{rev}}$ with $N = 10^6$ points. The time resolution was chosen as
\begin{equation}
	\Delta t = \frac{2\tau_\text{rev}}{N}, \qquad f_\text{s} = \frac{1}{\Delta t},
\end{equation}
where $f_\text{s}$ is the sampling frequency. The FFT\footnote{FFT stands for \emph{Fast Fourier Transform}, a computational algorithm used to efficiently evaluate the discrete Fourier transform (DFT) of a signal, converting it from the time domain into its frequency components.} of the sampled current signal yields a complex-valued spectrum $Y(f)$ over $N$ points. We define the two-sided amplitude spectrum by
\begin{equation}
	P_2(f) = \frac{1}{N} \big| \texttt{FFT} [I(t)]\big|,
\end{equation}
and the one-sided amplitude spectrum for positive frequencies as
\begin{equation}
	P_1(f) = \begin{cases}
		P_2(f), & f = 0, \\
		2P_2(f), & f > 0.
	\end{cases}
\end{equation}
The frequency axis is given by
\begin{equation}
	f_n = \frac{n f_\text{s}}{N}, \qquad n = 0, \ldots, N/2.
\end{equation}
To assess consistency with the analytical model, we normalize both $P_1(f)$ and the envelope $\tilde{I}_\text{b}(f)$ to their respective maxima and overlay them.
\begin{figure}[t]
	\centering
	\includegraphics[ width= \columnwidth]{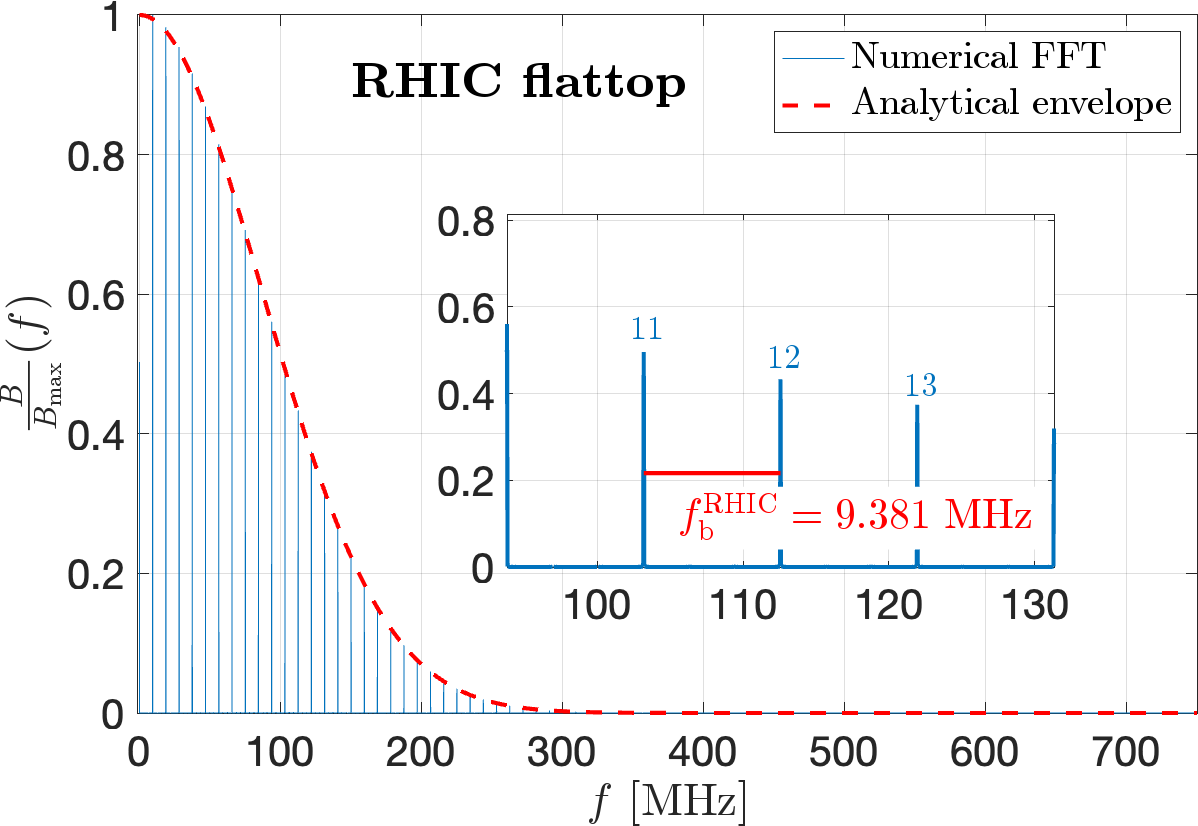}
	\caption{Comparison of the numerically obtained one-sided normalized FFT amplitude spectrum (blue) of the RF magnetic field $B(f)$ with the analytical envelope (dashed red) from Eq.\,\eqref{eq:gaussian-ft} for the conditions on RHIC flattop. The frequency axis is shown in MHz. Harmonic numbers $n = f / f_\text{b}$ are labeled near the peaks. The numerically computed bunch repetition frequency $f_\text{b}^\text{RHIC}$ in the inset agrees well with the analytically calculated one from Table\,\ref{tab:bunch-params}. }
	\label{fig:fft-comparison-RHIC}
\end{figure}

Figure\,\ref{fig:fft-comparison-RHIC} confirms that the numerical FFT closely follows the analytic envelope $\tilde{I}_\text{b}(f)$ over more than an order of magnitude. The few peaks shown in the inset appear at integer multiples of $f_\text{b}$, labeled by their harmonic number $n = f / f_\text{b}$, as expected from the periodic pulse structure. The FFT result shown in Fig.\,\ref{fig:fft-comparison-RHIC} is proportional to the spectral amplitude of the RF magnetic field $B(f)$ generated by the bunched beam at the target. The $y$-axis is labeled as $B/B_{\rm max}$ to reflect the normalization. For depolarization processes, however, the number of RF photons is proportional to the field power $|B(f)|^2$, i.e., the square of the displayed quantity.

\subsubsection*{Resolution limit of the discrete spectrum}

The frequency resolution $\Delta f = f_\text{s}/N$ in this analysis is governed by the total time window $T = N \Delta t$, so that
\begin{equation}
	\Delta f = \frac{1}{T} = 39.1\text{~kHz}
\end{equation}
for the chosen parameters, providing approximately 240 frequency bins per harmonic spacing of $f_b$ and adequate resolution to identify the resonance conditions within $\pm 19$~kHz required for hyperfine transition analysis. Sufficient spectral resolution requires a long sampling interval in time, whereas frequency coverage is determined by the sampling rate $f_\text{s}$.

\section{Beam-induced depolarization of hydrogen at RHIC}
\label{sec:rhic-depol}

We now examine how the RF spectrum of the circulating RHIC beam interacts with the internal hyperfine structure of hydrogen atoms in the target. The analysis evaluates resonance conditions, calculates photon emission rates, determines spatial field distributions, and assesses the impact on target polarization to establish operational safety thresholds.

The RHIC flattop conditions analyzed in this section serve to develop and validate the computational framework, which is subsequently applied to EIC  injection and  flattop scenarios in Section\,\ref{sec:EIC-depol}.

\subsection{Hyperfine transitions and resonance conditions}
\label{sec:hyperfine-transitions}

The bunched proton beam at RHIC generates a broadband spectrum of time-varying electromagnetic fields that can resonantly drive transitions between hyperfine levels in hydrogen atoms. These transitions are induced primarily by the magnetic component of the beam's RF field, which couples to the magnetic dipole moments of the atom.

The depolarization of atomic hydrogen in the presence of the RHIC beam arises when the frequency of a beam-induced RF magnetic field matches a hyperfine transition frequency $f_{ij}(B)$ at a given holding field $B$. Since the beam spectrum consists of discrete harmonics of the bunch frequency $f_\text{b} \approx \SI{9.381}{MHz}$, resonant transitions are possible when
\begin{equation}
	f_{ij}(B) = n \cdot f_\text{b}, \quad n \in \mathbb{N} \,.
	\label{eq:harmonic-match}
\end{equation}

Figure \ref{fig:transitions} shows the field dependence of the six hyperfine transition frequencies in absolute units (GHz). These cover a range from below \SI{0.1}{GHz} up to \SI{6}{GHz} as $B$ varies from 0 to \SI{200}{mT}. Not all six hyperfine transitions shown in Fig.\,\ref{fig:transitions} contribute to depolarization. Transitions that leave the nuclear spin quantum number $m_I$ unchanged, such as $|1\rangle \leftrightarrow |4\rangle$ and $|2\rangle \leftrightarrow |3\rangle$,  do not affect the hydrogen nuclear polarization in the target and are therefore excluded from further analysis. However, when analyzing the polarization of the ensemble using the BRP, the transitions between states with the same nuclear spin must be considered, as they affect the transmission through the sextupole magnets, and thus the polarization measurement in the BRP.

\begin{figure}[b]
	\centering
	\includegraphics[width= \columnwidth]{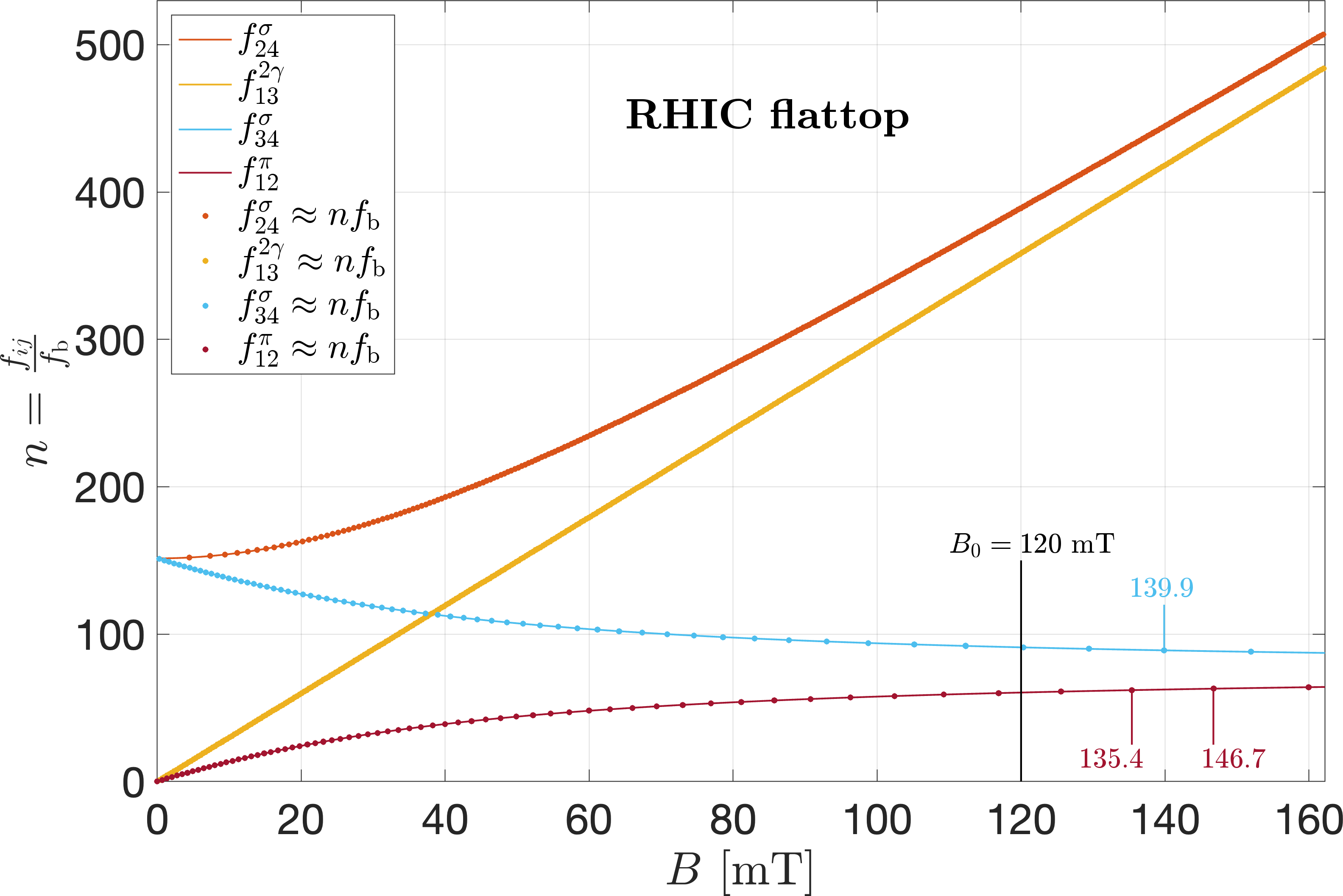}
	\caption{Hyperfine transition frequencies $f_{ij}(B)$ expressed as harmonic numbers $f_{ij}(B)/f_\text{b}$, relevant for the RHIC bunch structure (Fig.\,\ref{fig:bunch-profiles}). Dots indicate resonance points where the transition frequency satisfies $f_{ij}(B) \approx n f_\mathrm{b}$ within a tolerance of 0.002, corresponding to harmonic overlap with the bunch spectrum. In the vicinity of the static magnetic holding field $B_0 = \SI{120}{mT}$, the spacing between adjacent relevant resonance points is approximately \SI{4}{mT}.}
	\label{fig:hfs-transitions-harmonics-RHIC}
\end{figure}

To identify potential depolarization resonances, we evaluate the magnetic-field dependence of the remaining four transitions and express them both in absolute units (GHz) and in terms of the harmonic number $n = f_{ij} / f_\text{b}$, relative to the RHIC bunch frequency $f_\text{b} \approx \SI{9.381}{MHz}$. The visualization in Fig.\,\ref{fig:hfs-transitions-harmonics-RHIC}  illustrates where resonant conditions are met. For example, at the magnetic field of $B_0 \approx \SI{120}{mT}$ where the hydrogen jet target is operated, multiple transitions such as $f_{12}^\pi$, $f_{13}^{2\gamma}$, and $f_{34}^\sigma$ lie within a few MHz of a beam harmonic. Such coincidences open depolarization channels, provided the RF spectral power at the corresponding harmonic is sufficiently large. 

To visualize these resonant conditions, the harmonic number $n = f_{ij}(B) / f_\text{b}$ is plotted as a function of $B$ for each relevant hyperfine transition. Discrete markers highlight those magnetic field values where the transition frequency closely matches an integer multiple of the bunch frequency, specifically when
\begin{equation}
	\left| \frac{f_{ij}(B)}{f_\text{b}} - m \right| < 0.002\,, \quad \text{with } m \in \mathbb{Z}\,.
\end{equation}
These resonance conditions establish which hyperfine transitions can potentially be driven by the beam spectrum, but do not determine whether sufficient RF power exists at those frequencies to cause significant depolarization.

\subsection{Photon emission rate and spectral thresholds}
\label{sec:photon-emission-RHIC}

Having identified the resonance conditions for hyperfine transitions, we now estimate whether the beam-induced RF field carries sufficient power at those frequencies to induce significant depolarization.

\subsubsection{Theoretical framework and broadening effects}
\label{sec:theoretical-framework-power=broadening}

The frequency-domain envelope of the bunch train is governed by the Fourier transform of the single-bunch Gaussian profile, given in Eq.\,(\ref{eq:gaussian-ft}). This describes the spectral amplitude $\tilde{I}_\text{b}(f)$ in terms of the peak bunch current $I_\text{b}^\text{pk}$ and the RMS bunch width $\sigma_t$, and determines the harmonic content of the RF fields generated by the circulating beam.

To convert this current spectrum into a magnetic field amplitude spectrum $B(f)$ at a transverse distance $r$ from the beam axis, we use the expression
\begin{equation}
	B(f) = \frac{\mu_0}{2\pi r} \cdot I(f)\,,
	\label{eq:Boff}
\end{equation}
where $\mu_0 = 4\pi \times \SI{e-7}{H/m}$ is the permeability of free space. The expression for $B(f)$ follows from the Biot-Savart law for a straight current element at distance $r$ from the beam axis.

\begin{figure}[t]
	\centering
	\includegraphics[width =  0.9\columnwidth]{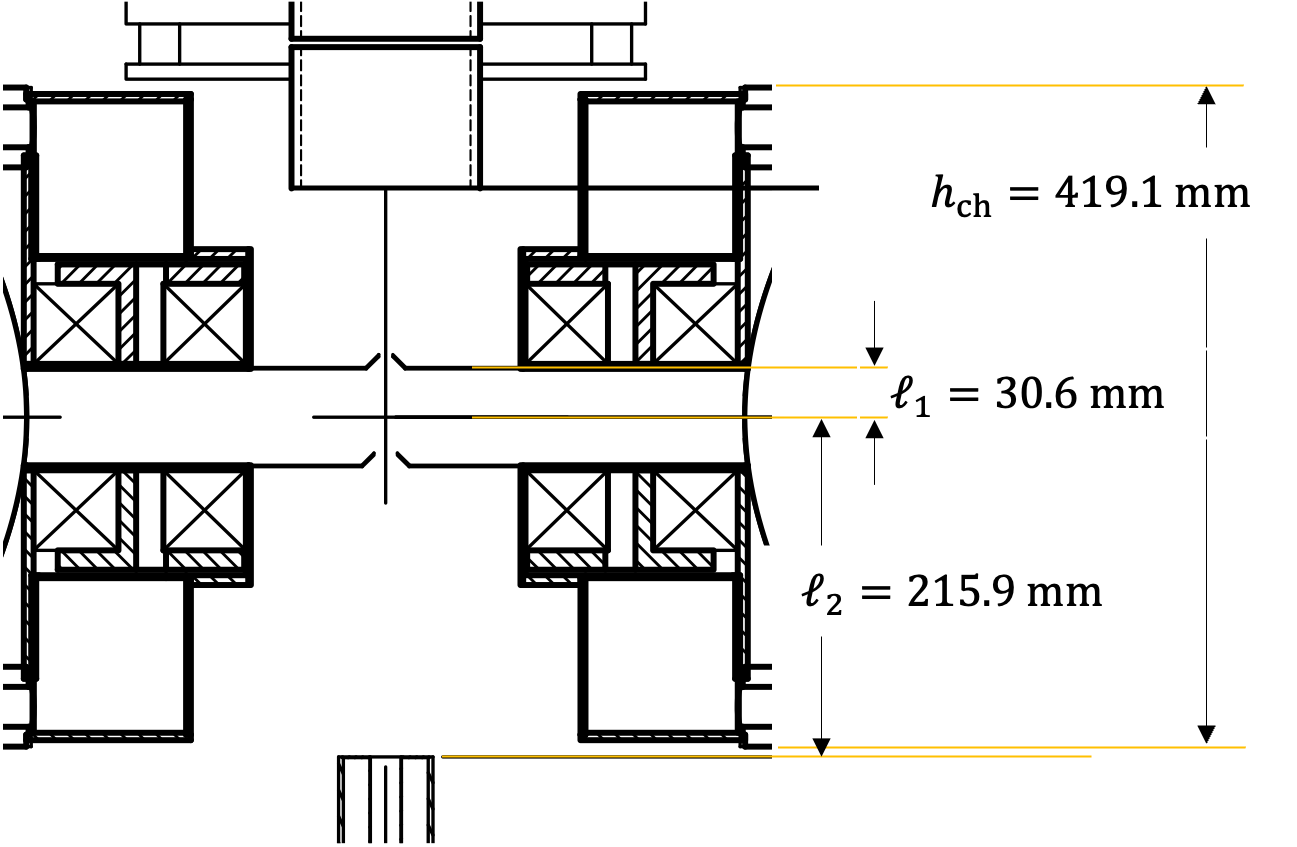}
	\caption{Side view of the RHIC target chamber to illustrate the interaction volume (see also Fig.\,\ref{fig:HJET}). The atomic beam enters from the top. The height of the target chamber is $h_\text{ch} = \SI{419.1}{mm}$. The distance between the exit of the RF transition unit in the ABS and the RHIC beam amounts to $\ell_1 = \SI{30.6}{mm}$.  The beam is assumed to have a transverse radius of $\approx \SI{5}{mm}$ as it travels downwards  into the BRP.}
	\label{fig:interaction-volume}
\end{figure}

The energy density associated with the magnetic field amplitude at frequency~$f$ is given by
\begin{equation}
	u(f) = \frac{B(f)^2}{\mu_0},
\end{equation}
so that the photon emission rate per unit bandwidth becomes
\begin{equation}
	\dot{N}_\gamma(f) = \frac{u(f)}{h f} \frac{V_\text{int}}{\tau_\text{int}} = \frac{1}{\mu_0}\frac{B(f)^2}{h f} \frac{V_\text{int}}{\tau_\text{int}}\,.
	\label{eq:photonrate}
\end{equation}
Here $V_\text{int} = L_{\text{int}} \cdot \pi r_{\text{at.\,beam}}^2 \approx \SI{2.40e-6}{m^3}$ is the effective interaction volume swept out by the atomic beam of radius $r_{\text{at.\,beam}} = \SI{5}{mm}$ along the interaction length $L_{\text{int}} = \ell_1 = \SI{30.6}{mm}$ in the upper half of the RHIC target chamber (see Fig.\,\ref{fig:interaction-volume}). The interaction time is $\tau_\text{int} =  L_{\text{int}} / v_\text{atom}$ and the atomic beam velocity $v_\text{atom} \approx \SI{1807}{m/s}$\,\cite{WISE20061}, yielding $\tau_\text{int} \approx \SI{17}{\micro\second}$. 

In this approximation, we neglect both the velocity distribution of the atomic beam and the finite width of the hyperfine resonances, treating transitions as occurring at discrete harmonic frequencies with a single representative atomic velocity $v_\text{atom}$.

However, these previously neglected effects introduce significant broadening mechanisms that influence the resonance conditions. For hydrogen atoms  emitted from a thermal source at temperature $T = \SI{80}{\kelvin}$, the Maxwell-Boltzmann velocity distribution
\begin{equation}
	f(v) = \sqrt{\frac{2}{\pi}} \left( \frac{m_{\text{H}}}{k_B T} \right)^{3/2} v^2 \exp\left( -\frac{m_{\text{H}} v^2}{2 k_B T} \right)\,,
\end{equation}
where $m_\text{H}$ is the mass of the hydrogen atom, and $k_B$ the Boltzmann constant (Table\,\ref{tab:Bc-constants}) which yields a thermal velocity spread along the beam axis with standard deviation
\begin{equation}
	\sigma_{\text{thermal}} = \sqrt{\frac{k_B T}{m_{\text{H}}}} = \SI{812}{m/s}\,.
\end{equation}
This velocity distribution results in Doppler broadening of the transition frequency with standard deviation
\begin{equation}
	\sigma_f^\text{Doppler} = f_0 \cdot \frac{\sigma_{\text{thermal}}}{c} \approx f_0 \cdot 2.71 \times 10^{-6}.
\end{equation}
For the hyperfine transition at $f_0 = \SI{1.42}{\giga\hertz}$, this yields
\begin{equation}
	\sigma_f^\text{Doppler} \approx \SI{3.85}{\kilo\hertz}\,.
\end{equation}

Additionally, power broadening arises when the RF magnetic field induces magnetic moment precession at the Rabi frequency
\begin{equation}
	f_\text{Rabi} = \gamma_\text{H} B_1 = \frac{g_J \mu_B B_1}{2\pi \hbar},
\end{equation}
where $B_1 = \SI{200}{\micro\tesla}$ represents the RF field amplitude averaged over the frequency spectrum (approximately $3 \sigma$ of the Gaussian envelope shown in Fig.\,\ref{fig:fft-comparison-RHIC}), accounting for the range of frequencies that contribute to power broadening and $\gamma_\text{H} / 2\pi \approx \SI{28.025}{\giga\hertz\per\tesla}$ is the gyromagnetic ratio of the hydrogen ground state (Table\,\ref{tab:Bc-constants}). This yields a precession frequency of
\begin{equation}
	f_\text{Rabi} \approx \SI{5.61}{\mega\hertz}.
\end{equation}
For consistent treatment with the Doppler contribution, the effective power broadening is expressed as a standard deviation via $\sigma^\text{power}_f = f_\text{Rabi}/(2\sqrt{2\ln 2})$, so that the combined effective linewidth, assuming Gaussian contributions, is  given by
\begin{equation}
	\sigma^\text{total}_f = \sqrt{ \left(\sigma_f^\text{Doppler}\right)^2 + \left( \sigma_f^\text{power})\right)^2 } \approx \SI{2.38}{\mega\hertz}\,.
\end{equation}

This broadening has implications for the harmonic analysis and leads to a fundamental limitation of our approach. The discrete harmonic method identifies resonance conditions by requiring exact frequency matches between hyperfine transition frequencies and beam harmonic frequencies.  Since our analysis only flags exact frequency matches, it provides a lower bound on depolarization risks by not accounting for these near-resonant effects.

The implications for the harmonic spacing are favorable: while the \SI{2.38}{\mega\hertz} linewidth is much larger than the precision required for exact matching, it remains small compared to the harmonic spacing (\SI{9.381}{\mega\hertz}), ensuring that neighboring harmonics do not overlap. This validates the discrete harmonic approach while acknowledging that additional transitions within \SI{2.38}{\mega\hertz} of any harmonic frequency could exhibit resonant behavior beyond what our threshold determination captures.

\subsubsection{Quantitative analysis and threshold determination}

We now apply the theoretical framework developed above to calculate the actual photon emission rates and determine depolarization thresholds for RHIC operating conditions.

To account for the vertical variation of the azimuthal magnetic field along the atomic flight path in the upper half of the chamber, the field amplitude $B(f)$ in Eq.\,(\ref{eq:photonrate}) is replaced by its vertical average $\langle B(f) \rangle$, defined as
\begin{equation}
	\langle B(f) \rangle = \frac{1}{L_{\text{int}}} \int_0^{L_{\text{int}}} B(f, r) \, \dd r \,,
	\label{eq:Bavg_vertical}
\end{equation}
so that $B(f)^2 \to \langle B(f) \rangle^2$ in Eq.\,(\ref{eq:photonrate}). This averaging is necessary because atoms travel through regions of varying magnetic field strength along their vertical flight path toward the target region (see Fig.\,\ref{fig:interaction-volume}).

To obtain the total time-averaged photon flux of the full circulating beam from Eq.\,(\ref{eq:photonrate}), the spectral emission rate must be scaled by the effective duty cycle. Defining the average photon emission rate as $\dot{N}_\gamma^{\mathrm{avg}}(f)$, we write
\begin{equation}
	\dot{N}_\gamma^{\mathrm{avg}}(f) = \dot{N}_\gamma(f) \cdot f_\text{b} \cdot \tau_t\,,
	\label{eq:dotN_gamma_avg}
\end{equation}
where $f_\text{b}$ is the bunch repetition frequency (Table\,\ref{tab:bunch-params}) and $\tau_\text{t} = 2\sqrt{2\ln 2} \cdot \sigma_t \approx \SI{4.32}{ns}$ is the FWHM of the temporal bunch duration. This correction reflects the fact that significant magnetic field amplitudes exist only during the brief bunch passage. The result, $\dot{N}_\gamma^{\mathrm{avg}}(f)$, represents the physically relevant time-averaged spectral photon rate.

\begin{figure}[tb]
	\centering
	\includegraphics[width= \columnwidth]{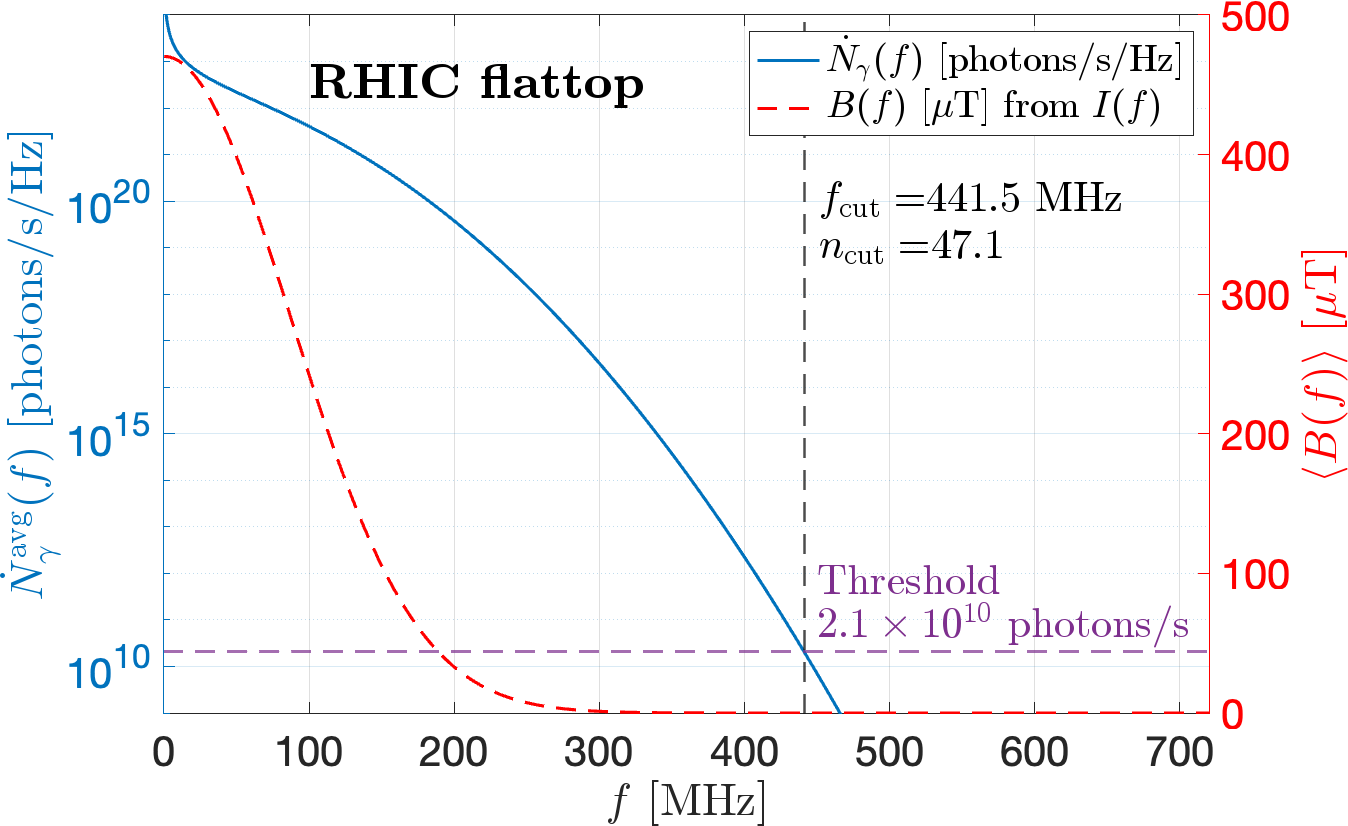}
	\caption{Photon emission rate $\dot{N}^\text{avg}_\gamma(f)$ from Eq.\,(\ref{eq:dotN_gamma_avg}) (left axis, blue line) and corresponding RF magnetic field amplitude $B(f)$ (right axis, dashed line), both derived from the analytical Gaussian RHIC bunch envelope in Eq.\,(\ref{eq:gaussian-ft}) and converted using Eq.\,(\ref{eq:Boff}). The photon rate is computed from the energy density using Eq.\,(\ref{eq:photonrate}). The dashed horizontal line marks the threshold of $\SI{2.1e10}{photons/s/Hz}$ required to depolarize about 1\% of the atoms in the beam. The vertical line marks the cutoff frequency $f_{\text{cut}}$ and harmonic number $n_{\text{cut}}$ where the photon rate drops below the depolarization threshold.}
	\label{fig:photon-vs-field-spectrum-RHIC}
\end{figure}

The result of this calculation is shown in Fig.\,\ref{fig:photon-vs-field-spectrum-RHIC}, where the left axis displays the photon emission rate $\dot{N}^\text{avg}_\gamma(f)$, and the right axis shows the corresponding magnetic field amplitude $B(f)$. To induce significant depolarization, the photon emission rate at a given harmonic must be high enough to affect a non-negligible fraction of atoms present in the interaction volume at any given moment. Based on typical HJET operating conditions, the atomic flux through the interaction region is $\Phi = (12.4 \pm 0.2) \times \num{e16}\,\si{atoms\per s}$ with a jet target thickness along the RHIC beam of $(1.3 \pm 0.2) \times \num{e12}\, \si{atoms \per  cm \squared}$\,\cite{Zelenski2005}. Given the atomic flux $\Phi$ and a beam transit time through the interaction region of $\tau_\text{int}$ from above, the instantaneous number of atoms in the chamber is
\begin{equation}
	N_\text{atoms} = \Phi \cdot \tau_\text{int} \approx 2.1 \times 10^{12}\,.
\end{equation}
To achieve 1\% depolarization, representing a measurable change that would significantly impact the nuclear target polarization and exceed the required systematic uncertainties, a photon rate of at least $2.1 \times \num{e10}\,\si{photons/s/Hz}$ is required at resonance. This value sets a threshold, which is shown as a reference line in Fig.~\ref{fig:photon-vs-field-spectrum-RHIC}.

For RHIC flattop, the intersection point where the photon emission rate $\dot{N}^\text{avg}_\gamma(f)$ drops below the threshold occurs at a frequency $f_{\mathrm{cut}} \approx \SI{441.5}{MHz}$, corresponding to harmonic number $n_{\mathrm{cut}} \approx 47$. Above this frequency, the photon flux is insufficient to depolarize a significant fraction of the atomic beam, making higher harmonics increasingly ineffective. However, this estimate involves uncertainties: unfortunately, no dedicated polarization measurements with the BRP and varying magnetic holding field have been performed at RHIC with stored beam to locate the true depolarization onset, and transient beam-induced fields may locally shift atoms into resonance. 
The following section provides a quantitative estimate of the relevant magnetic fields in the interaction region.

\subsection{Instantaneous magnetic field at the target}
\label{sec:Bfield-spatial-RHIC}

We now quantify the instantaneous magnetic field generated by the beam  bunch as it passes the atomic target, based on the spatial current distribution of the beam.

To estimate the magnetic field amplitude experienced by atoms in the target due to the circulating beam, we model the transverse distribution of a single bunch as a two-di\-men\-sion\-al Gaussian
\begin{equation}
	\rho(x, y) = \frac{1}{2\pi \sigma_x \sigma_y} \exp\left(-\frac{x^2}{2\sigma_x^2} - \frac{y^2}{2\sigma_y^2}\right)\,,
	\label{eq:rho_gaussian}
\end{equation}
where $\sigma_{x,y}$ are the horizontal and vertical RMS beam sizes at the interaction point. This expression allows for asymmetric (elliptical) beams; the round-beam case corresponds to $\sigma_x = \sigma_y \equiv \sigma_r$.

Assuming that the longitudinal and transverse distributions factorize and the beam propagates along the $z$-axis, the current density becomes
\begin{equation}
	\vec{J}(x, y, z, t) = \vec e_z \cdot I_\text{b}(t) \cdot \rho(x, y),
\end{equation}
where $I_\text{b}(t)$ is the time-dependent longitudinal bunch current profile, defined in Eq.\,(\ref{eq:gaussian-bunch}) with peak current $I^\text{pk}_\text{b}$ from Table\,\ref{tab:bunch-params}.

\subsubsection*{Round beam profiles}

We distinguish between round and elliptic transverse beam profiles to evaluate how the bunch geometry influences the spatial dependence of the magnetic field at the target.

The magnetic field at a transverse point $\vec{r} = (x, y)$ (e.g., where an atom in the target is located) is obtained from the Biot-Savart law,
\begin{equation}
	\vec{B}(\vec{r}, t) = \frac{\mu_0}{4\pi} \int \frac{\vec{J}(\vec{r}\,') \times (\vec{r} - \vec{r}\,')}{|\vec{r} - \vec{r}\,'|^3} \, \dd^3r',
	\label{eq:biot-savart}
\end{equation}
which yields a magnetic field $\vec{B} = B(r, t)\, \vec e_{\phi}$, oriented in the azimuthal direction $\vec{e}_\phi$, which is defined by the right-hand rule as 
$\vec{e}_\phi = \vec{e}_z \times \vec{e}_r$. This results in
\begin{equation}
	\vec{B}(r, t) = \frac{\mu_0}{2\pi r} \cdot I_\text{b}(t) \cdot F(r) \, \vec{e}_\phi \,,
	\label{eq:B_r_with_F}
\end{equation}
where $F(r)$ is a dimensionless geometric correction factor that accounts for the spatial extension of the transverse beam distribution. For a round Gaussian beam, $F(r)$ can be evaluated analytically via
\begin{equation}
	F(r) = 1 - \exp\left(-\frac{r^2}{2\sigma_r^2}\right),
	\label{eq:F_r}
\end{equation}
with $\sigma_r = \sigma_x = \sigma_y$. In the limit $r \gg \sigma_r$, the expression reduces to the standard Biot-Savart result for a line current,
\begin{equation}
	B(r, t) \approx \frac{\mu_0}{2\pi r} \cdot I_\text{b}(t).
\end{equation}

To analyze the spectral content, we take the Fourier transform of the time-dependent current profile,
\begin{equation}
	B(f, r) = \frac{\mu_0}{2\pi r} \cdot I(f) \cdot F(r),
	\label{eq:Bf_full}
\end{equation}
where $I(f)$ is the current amplitude spectrum defined in Eq.\,(\ref{eq:gaussian-ft}).

\subsubsection*{Elliptic beam profiles}
\label{sec:elliptic-beam}

In the general case where $\sigma_x \neq \sigma_y$, the beam has an elliptical transverse profile. The Biot-Savart integral in Eq.~\eqref{eq:biot-savart} must be evaluated numerically for arbitrary field points $\vec{r}$. To handle this more complex geometry efficiently, we employ a vector potential approach.

The magnetic field $\vec{B}(\vec{r})$ generated by a steady current distribution $\vec{J}(\vec{r}\,')$ can be expressed using the vector potential formalism,
\begin{equation}
	\vec{B}(\vec{r}) = \nabla \times \vec{A}(\vec{r})\,,
\end{equation}
where the vector potential $\vec{A}(\vec{r})$ satisfies the Poisson equation
\begin{equation}
	\nabla^2 \vec{A}(\vec{r}) = -\mu_0 \vec{J}(\vec{r})\,.
\end{equation}
For a current flowing in the $z$-direction with a 2D Gaussian transverse profile, the vector potential has only a $z$-component. Using the appropriate Green's function for the 2D Laplacian, this component can be expressed as
\begin{equation}
	A_z(\vec{r}) = -\frac{\mu_0}{2\pi} \iint J_z(\vec{r}\,') \ln\frac{1}{|\vec{r} - \vec{r}\,'|} \, dS',
\end{equation}
where $J_z(\vec{r}\,')$ is the current density distribution for the elliptical Gaussian beam,
\begin{equation}
	J_z(x',y') = I_\text{b} \cdot \rho(x',y')\,,
\end{equation}
with $\rho(x,y)$ as defined in Eq.~\eqref{eq:rho_gaussian}.

The magnetic field components are then obtained from the curl of $\vec{A}$ via
\begin{equation}
	B_x = \frac{\partial A_z}{\partial y}, \quad 
	B_y = -\frac{\partial A_z}{\partial x}, \quad
	B_z = 0\,.
\end{equation}
Since the vector potential has only a $z$-component and we are examining the 2D transverse Gaussian current distribution at a fixed instant (at the peak of the bunch), the magnetic field at this moment has no longitudinal component ($B_z = 0$).

This vector potential approach inherently handles the potential singularity in the Biot-Savart law through the naturally regularizing properties of the Gaussian current distribution, while enabling efficient numerical implementation on a discrete grid. Unlike the round beam case, the resulting magnetic field becomes direction-dependent even at fixed radial distance, making this treatment essential for the elliptical beam profiles expected at the location of the polarized target in IP4 at the EIC.

\subsection{Spatial field distribution}

We now turn to the spatial profile of the peak magnetic field amplitudes at the target, emphasizing their dependence on beam optics parameters such as emittance and beta function.

To evaluate the magnetic field amplitude $B(f, r)$ experienced at a given transverse offset $r$, we require knowledge of the transverse beam dimensions $\sigma_{x}$ and $\sigma_{y}$. For RHIC, these are derived from the normalized emittance $\epsilon_\text{n}$ and local beta functions $\beta_{x,y}$ at the present  target location at IP12. The transverse RMS beam sizes are given by
\begin{equation}
	\sigma_{x,y} = \sqrt{\frac{\beta_{x,y} \, \epsilon^\text{n}_{x,y}}{\beta \gamma}}\,,
	\label{eq:sigma-beam}
\end{equation}
where  $\beta$ and $\gamma$ are the relativistic factors. 
To convert from RMS to 95\% normalized emittance, a factor of 5.993 is used in one dimension, as discussed in Ref.~\cite{LeeAcceleratorPhysics2011}, so that 
\begin{equation}
	\begin{split}
		 \epsilon^\text{n, 95}_{x,y} & = \epsilon_{x,y}^\text{n} \cdot 5.933\,, \, \text{and} \\
		 \sigma_{x,y}^{95} & = \sigma_{x,y} \cdot \sqrt{5.933}\,.
	\end{split}
\end{equation}

Table\,\ref{tab:bunch-params} summarizes the relevant beam and optics parameters at IP\,12 for RHIC at flattop ($E = \SI{255}{GeV}$). The normalized RMS emittance was taken from the RHIC dashboard during run 22.

\begin{figure}[hbt]
	\includegraphics[width= \columnwidth]{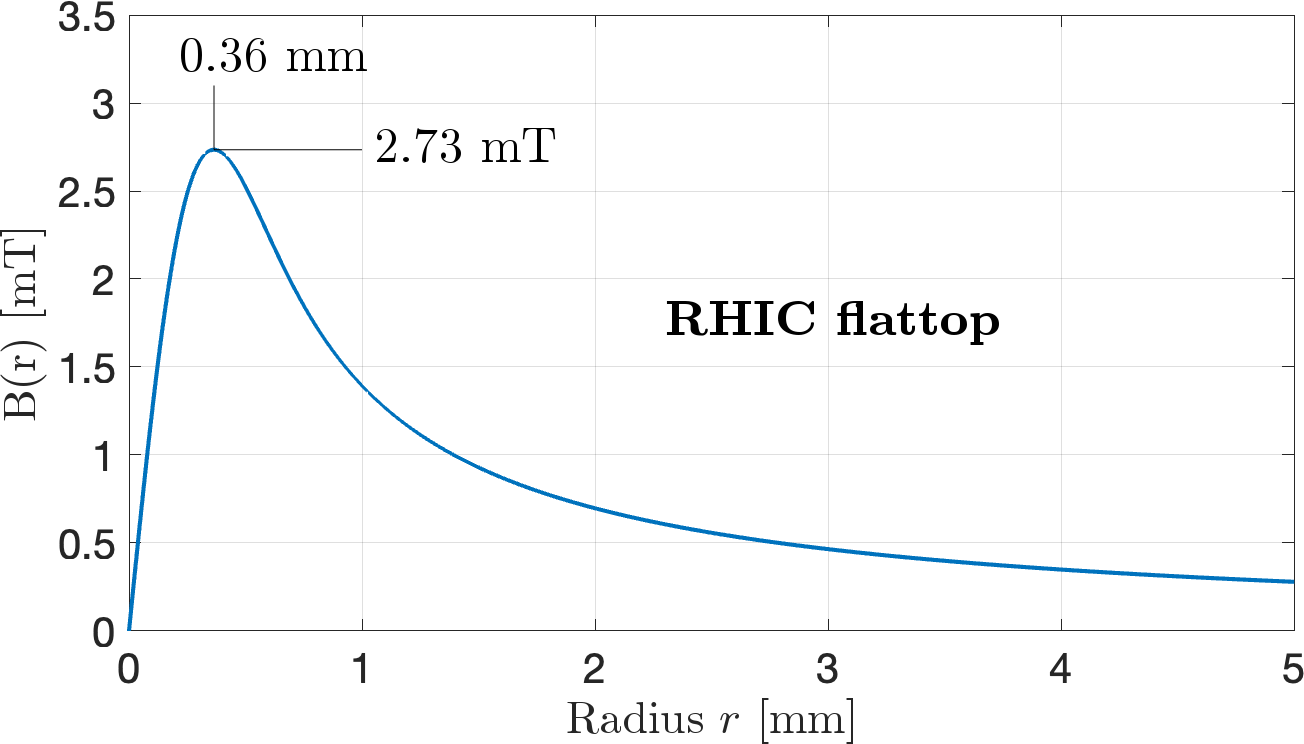}
	\caption{Peak azimuthal magnetic flux density B(r) produced by a single bunch at RHIC at flattop energy \SI{255}{GeV} as a function of transverse distance $r$ from the beam center. The field amplitude is evaluated at the peak current using  Eq.\,\eqref{eq:B_r_with_F}. Vertical and horizontal markers indicate the field maximum and its location.}
	\label{fig:Bfield-beam-RHIC}
\end{figure}

Figure\,\ref{fig:Bfield-beam-RHIC} shows the peak magnetic flux density $B(r)$ produced by a passing RHIC bunch as a function of transverse distance $r$ from the beam axis, assuming a round Gaussian beam with RMS width $\sigma_r$ determined by the beta function and normalized emittance at the HJET location. The curve shows the \SI{255}{GeV} flattop energy, evaluated at the peak of the bunch distribution ($t = 0$) from Eq.\,(\ref{eq:gaussian-bunch}). The field drops off approximately as $1/r$ for $r \gg \sigma_r$.

In the vicinity of the nominal holding field $B_0 = \SI{120}{mT}$, as shown in Fig.\,\ref{fig:hfs-transitions-harmonics-RHIC}, the spacing between consecutive resonances for the $f^\sigma_{34}$ and $f^\pi_{12}$  transitions is approximately \SI{4}{mT}. Variations in the holding field can shift the system in and out of resonance with beam harmonics, potentially modulating the nuclear depolarization rate of the target.

It is important to note that the target polarization observed in the detector system is determined from many bunches that sequentially intercept the target. Depolarization effects are strongest when the bunch center coincides with the target location, corresponding to the peak of the beam-induced magnetic field. This localized and transient interaction can alter the spin composition of the sample seen by the detectors that measure scattered protons to the left and right of the target. In contrast, the Breit-Rabi polarimeter (BRP) measures the time-averaged spin population of atoms exiting the target chamber and may not resolve short-lived depolarization effects occurring only during bunch passage. Note that this paper does not investigate potential beam-induced effects on the BRP measurement itself.

\subsection{Impact on target polarization}

Having established the spatial and spectral characteristics of the beam-induced magnetic field at RHIC flattop, we now assess its impact on the target polarization through its influence on hyperfine transition conditions.

The azimuthal time-dependent magnetic field $\vec B(r, t)$ from Eq.\,(\ref{eq:B_r_with_F}) generated by the passing beam bunch reaches amplitudes of several \si{mT} near the beam axis and varies rapidly across the transverse extension of the atomic beam. This field plays a central role in determining whether hyperfine transitions can be driven resonantly. As shown in Fig.\,\ref{fig:Bfield-beam-RHIC}, for a circulating RHIC beam at flattop energy of $\SI{255}{GeV}$, the peak magnetic field amplitude is $B_\text{max} = \SI{2.73}{mT}$, occurring at a radial distance $r = \SI{0.36}{mm}$ from the beam axis, well within the atomic beam diameter of approximately \SI{10}{mm}\,\cite{Zelenski2005}, and more importantly, well within the transverse target area sampled by the RHIC beam, for which $\sigma_r^{95\%} = \SI{0.56}{mm}$ (see Table\,\ref{tab:bunch-params}). This corresponds to the location where the transverse field profile peaks for a round Gaussian beam. The resulting time-dependent RF field must be considered when assessing the proximity of hyperfine transition frequencies to harmonic components in the beam spectrum. For comparison, at RHIC injection energy (\SI{23.5}{GeV}, $\gamma \approx 25.05$), the beam size scales as $\sigma_{x,y} \propto \sqrt{\beta_{x,y}\epsilon_n/(\beta\gamma)}$, so that the radius is approximately $\sqrt{10}$ times larger, substantially reducing the maximum magnetic field amplitudes to about $B_\text{max} \approx \SI{0.70}{mT}$ at $r \approx \SI{1.4}{mm}$.

Since the local magnetic field shifts the hyperfine energy levels, the \emph{resonance condition} for transitions, given in Eq.\,(\ref{eq:harmonic-match}), can be modified \emph{locally} by the presence of the beam-induced  magnetic field $\vec B(r, t)$, even if the static holding field $\vec B_0$ is uniform. As the bunch passes, atoms at different transverse positions experience different instantaneous total magnetic fields,
\begin{equation}
	\vec B_\text{eff}(r, t) = B_0 \cdot \vec e_y + B(r, t) \cdot \vec e_\phi\,,
\end{equation}
where $B(r, t)$ is the magnitude of the azimuthal magnetic field from Eq.\,(\ref{eq:B_r_with_F}) and $B_0$ the static holding field from Eq.\,(\ref{eq:B0}). This superposition of static holding and beam-induced field alters not only the resonance condition for transitions but also the local magnitude and orientation of the magnetic field that defines the spin quantization axis of the nuclear target polarization. As a result, atoms on opposite sides of the beam axis experience different magnetic fields during the bunch passage. Since the hyperfine energy levels -- and thus the equilibrium nuclear polarization -- depend non-linearly on the local field strength, these field asymmetries induce small, spatially dependent variations in the target polarization. When the beam-target interaction is perfectly symmetric, these effects average out, but any asymmetry in the beam-target overlap (beam not perfectly centered, etc.) can lead to a net modification of the measured target polarization.

Averaging the azimuthal magnetic field across the beam radius out to $\sigma_r^{95}$ in the midplane ($y = 0$) yields a net offset of approximately $\SI{2.09}{mT}$. This breaks the left-right symmetry, since the effective average field becomes $B_{\rm L} = \SI{122.09}{mT}$ in the left hemisphere and $B_{\rm R} = \SI{117.91}{mT}$ on the right. This spatial variation leads to an imbalance in the nuclear polarization of atoms through which the stored beam passes. To quantify the effect for two injected states like $|1\rangle + |4\rangle$, we calculate the resulting difference in target polarization between the hemispheres through
\begin{equation}
	\label{eq:rel_pol_asymmetry-RHIC}
	\delta Q = \frac{Q_{|1\rangle+|4\rangle}(B_{\rm L}) - Q_{|1\rangle+|4\rangle}(B_{\rm R})}{Q_{|1\rangle+|4\rangle}(B_y^\text{nom})} \approx 0.25\%,
\end{equation}
where we have used the expressions from Eq.\,\eqref{eq:Q-of-x-for1to4}. The result is the same for states $|2\rangle + |3\rangle$, and the effect appears to be small for HJET operation at RHIC and does not contribute significant systematic uncertainty to the measured jet polarization.

\vspace{1em}

The analysis shows that under RHIC conditions, beam-in\-duced depolarization is unlikely to play a significant role. The time-averaged photon emission rate $\dot{N}\gamma^{\mathrm{avg}}(f)$ falls below the critical threshold of $2.1 \times 10^{10}$ photons/s/Hz above the cutoff frequency $f_\text{cut} \approx \SI{441.5}{MHz}$, corresponding to harmonic number $n_\text{cut} \approx 47$. To ensure robustness against local perturbations -- such as those from the beam's own transient magnetic fields -- it is prudent to treat $n_\text{cut}$ as a lower bound and avoid operation below a factor of $\approx 3$ of this limit. For comparison, Fig.\,\ref{fig:hfs-transitions-harmonics-RHIC} shows that RHIC flattop provides a safety factor of approximately 5 ($\approx 375/75$) for HJET operation. Furthermore, field-induced modifications to the effective holding field lead to a small target polarization imbalance across the atomic beam, with $\delta Q/Q \lesssim 0.2\%$ for the typical $|1\rangle + |4\rangle$ and $|2\rangle + |3\rangle$ injected state combinations. Overall, these results establish RHIC as a well-characterized reference point, providing the baseline for the EIC-specific evaluation in the next section.

\section{Beam-induced depolarization of hydrogen at the EIC}
\label{sec:EIC-depol}


Having established the computational framework using the RHIC conditions in Section\,\ref{sec:rhic-depol}, we now apply this methodology to evaluate beam-induced depolarization risks at the future EIC. The EIC presents new challenges due to higher bunch repetition frequencies, smaller beam sizes, and elliptical beam profiles. We assess depolarization risks for the operation of the polarized hydrogen target at EIC injection and flattop energies (\SI{23.5}{GeV} and  \SI{275}{GeV}). 

Unlike at RHIC, at injection, the hadron beams at EIC will undergo electron cooling for approximately \SI{30}{minutes} to reduce the vertical emittance, thereby providing an extended window for beam polarization calibration using the HJET. Measurements at both injection and flattop energies are essential to establish absolute polarization calibration points throughout the accelerator chain. With present-day polarized target technology and the anticipated hundreds to over a thousand bunches circulating in the EIC, these measurements will surpass both the systematic and statistical precision achievable in the Booster or AGS, where only single bunches or a few bunches can be stored. Furthermore, absolute polarization calibration is essential to understand polarization transmission through the accelerator chain, where for protons such calibration is currently only available at the \SI{200}{\mega\electronvolt} polarimeter behind the Linac\,\cite{Zelenski_2011}.

\subsection{EIC beam parameters, spectral characteristics, and depolarization thresholds for $B_0 = \SI{120}{mT}$}

In this section, we examine how the situation would appear if the polarized target were operated at the same $B_0 = \SI{120}{mT}$ holding field as at RHIC. 

The beam and optics parameters at the future HJET location in IP\,4 for both EIC energies are summarized in Table\,\ref{tab:bunch-params}. Compared to RHIC conditions, the EIC presents several key differences: significantly higher bunch repetition frequencies, smaller normalized emittances leading to reduced transverse beam sizes, and elliptical beam profiles due to unequal beta functions at the interaction point. These parameters alter the RF field strength, harmonic density, and spatial field distributions experienced by the hydrogen atoms, as it brings many more atomic transitions within the range of potentially depolarizing harmonics.

Importantly, the elliptical transverse beam profile at the EIC does not influence the frequency-domain spectrum, which depends solely on the longitudinal current distribution $I_\text{b}(t)$ and bunch spacing $f_\text{b}$. The beam-induced magnetic field spectrum $B(f)$ inherits this harmonic structure directly through Eq.\,\eqref{eq:Boff}, enabling direct application of the resonance analysis framework established in Section\,\ref{sec:rhic-depol}.

\begin{figure}[t]
	\centering
	\begin{subfigure}[b]{\columnwidth}
		\centering
		\includegraphics[width=\textwidth]{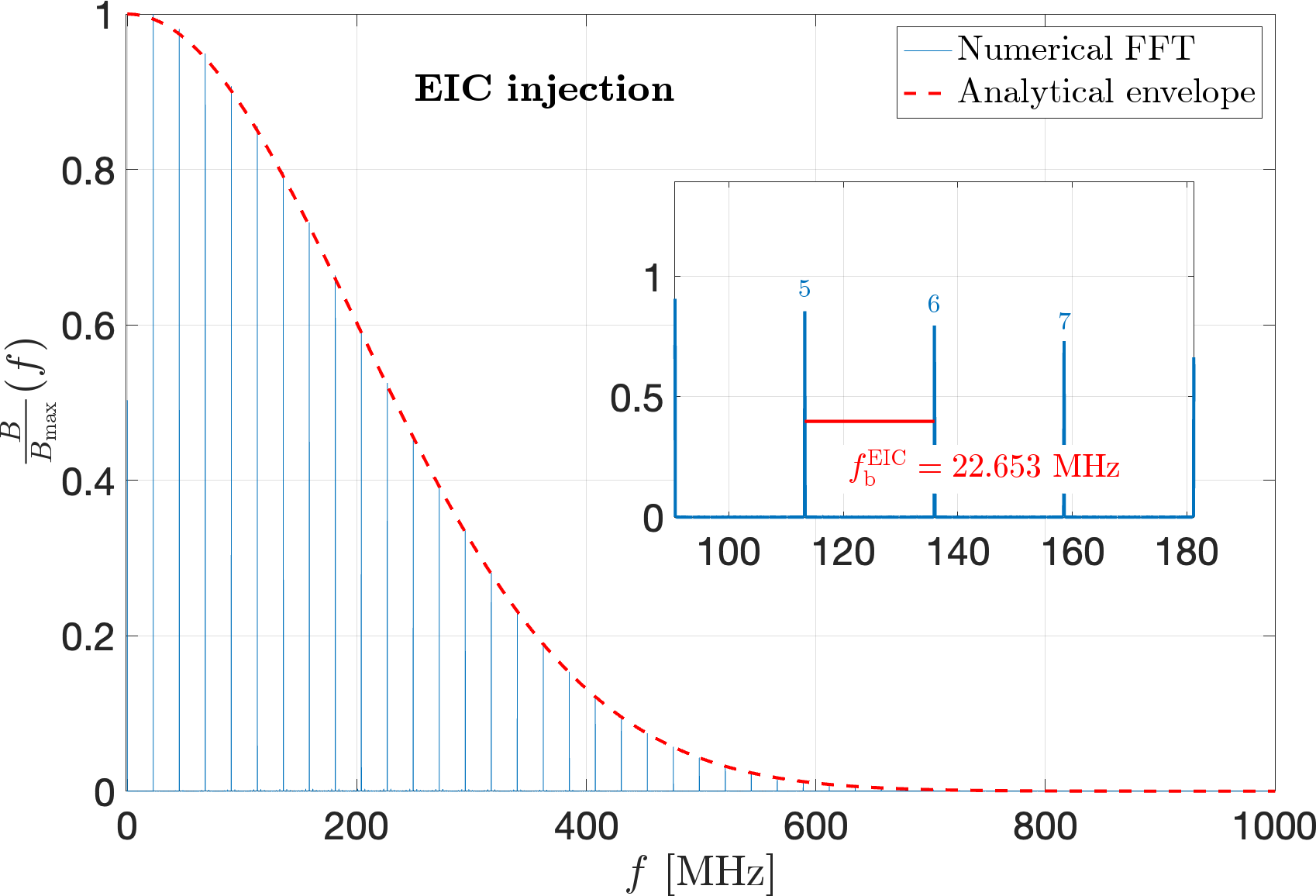}
		\caption{EIC injection (\SI{23.5}{GeV})}
		\label{fig:EIC-photon-rate-injection}
	\end{subfigure}
	\begin{subfigure}[b]{\columnwidth}
		\centering
		\includegraphics[width=\textwidth]{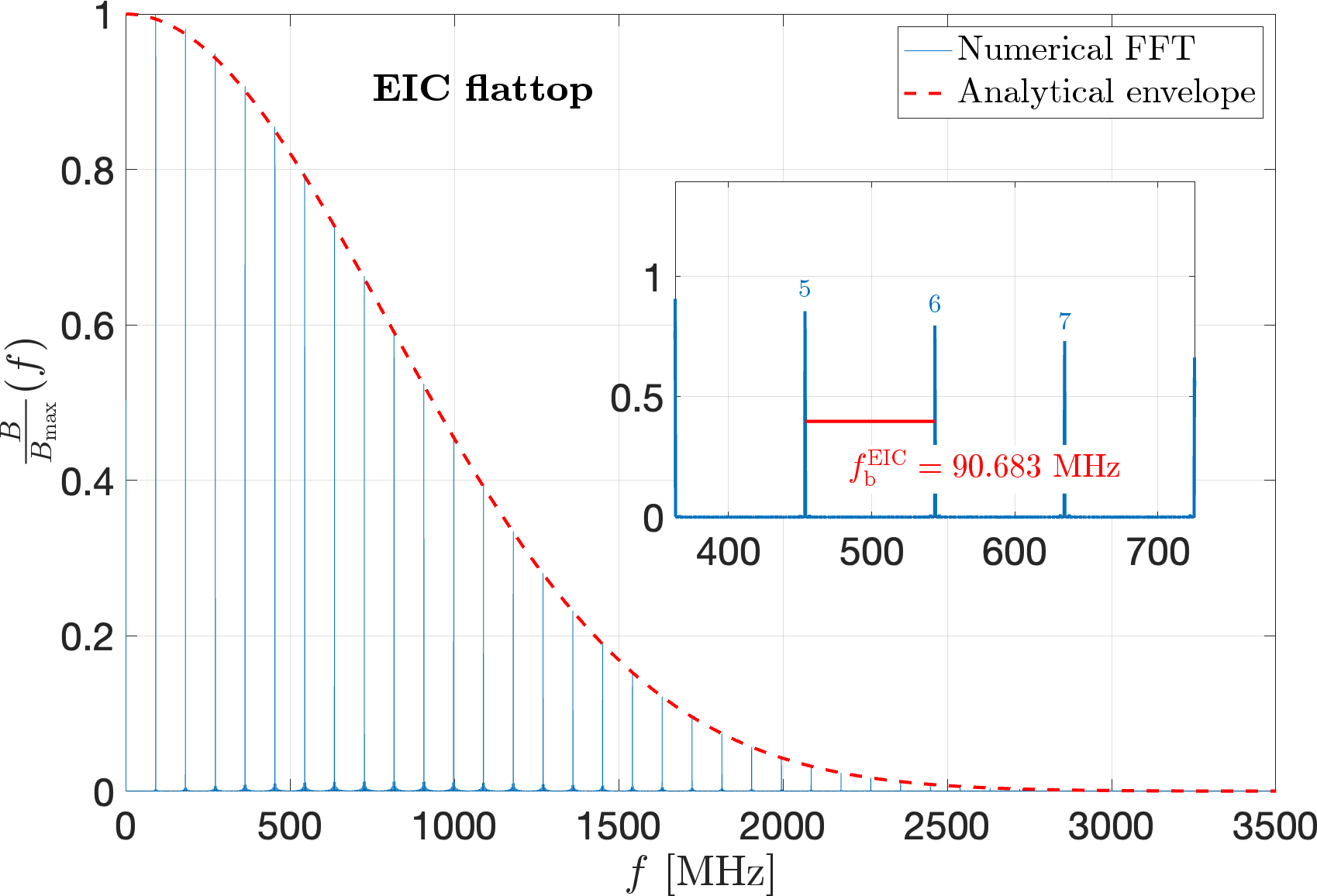}
		\caption{EIC flattop (\SI{275}{GeV})}
		\label{fig:EIC-photon-rate-flattop}
	\end{subfigure}
	\caption{Frequency-domain spectra for EIC bunch trains at (a) injection (23.5 GeV) and (b) flattop (275 GeV) energies. The plots show the numerically obtained one-sided normalized FFT amplitude spectrum (blue) overlaid with the analytical Gaussian envelope (dashed red) from Eq.\,\eqref{eq:gaussian-ft}, following the same methodology as Fig.\,\ref{fig:fft-comparison-RHIC}. The higher bunch repetition frequencies at EIC result in wider harmonic spacing compared to RHIC. Harmonic numbers $n = f/f_\text{b}$ are labeled for selected peaks.}
	\label{fig:EIC-photon-rate}
\end{figure}

The frequency-domain spectra of the EIC bunch trains at injection (23.5 GeV) and flattop (275 GeV) energies were numerically obtained alongside the analytical envelopes, in the same fashion as shown on Fig.\,\ref{fig:fft-comparison-RHIC}, making use of Eq.~(\ref{eq:gaussian-ft}) with the EIC-specific parameters from Table\,\ref{tab:bunch-params}, yielding a familiar  series of discrete harmonic peaks modulated by a Gaussian envelope. Compared to the RHIC spectrum (Fig.\,\ref{fig:fft-comparison-RHIC}), both EIC spectra shown in Fig.\,\ref{fig:EIC-photon-rate} indicate a considerably higher frequency content due to their shorter bunch durations and higher bunch frequencies. 

\begin{table}[t]
	\centering
	\caption{Result of the frequency-domain spectral analysis of the  bunch trains listing the obtained parameters $f_\text{cut}$ and harmonic cut off $n_\text{cut}$ required to depolarize about 1\% of the atoms in the atomic beam for RHIC and the two EIC cases (injection and flattop).}
	\label{tab:harmonic-cutoff-depol}
	\begin{ruledtabular}	
	\begin{tabular}{p{4.2cm}rrr}
		& RHIC & \multicolumn{2}{c}{EIC} \\ 			
		Quantity & flattop & injection & flattop \\ 	\hline
		bunch frequency $f_\text{b}$\,[\si{MHz}]  &  9.381 & 22.653 & 90.683  \\	
		cut-off frequency 
		$f_\text{cut}$\,[\si{MHz}] &  441.5 & 1039.1 & 4053.6\\
		harmonic cut off $n_\text{cut}(f_\text{cut})$ & 47.1 & 45.9 & 44.7 \\
		%
		%
	\end{tabular}
	\end{ruledtabular}
\end{table}

The photon emission rates  $\dot{N}^\text{avg}_\gamma(f)$ from Eq.\,(\ref{eq:dotN_gamma_avg}) for the two cases were analyzed to determine where the photon rate drops below the depolarization threshold in the same way as previously applied for RHIC in Fig.\,\ref{fig:photon-vs-field-spectrum-RHIC}. The cut-off frequency $f_\text{cut}$ and corresponding harmonic cut-off $n_\text{cut}$ were obtained to depolarize about 1\% of the atoms in the beam. The results are summarized in Table\,\ref{tab:harmonic-cutoff-depol}.

\begin{figure*}[hbt]
	\centering
	\begin{subfigure}[t]{\columnwidth}
		\centering
		\includegraphics[width=\columnwidth]{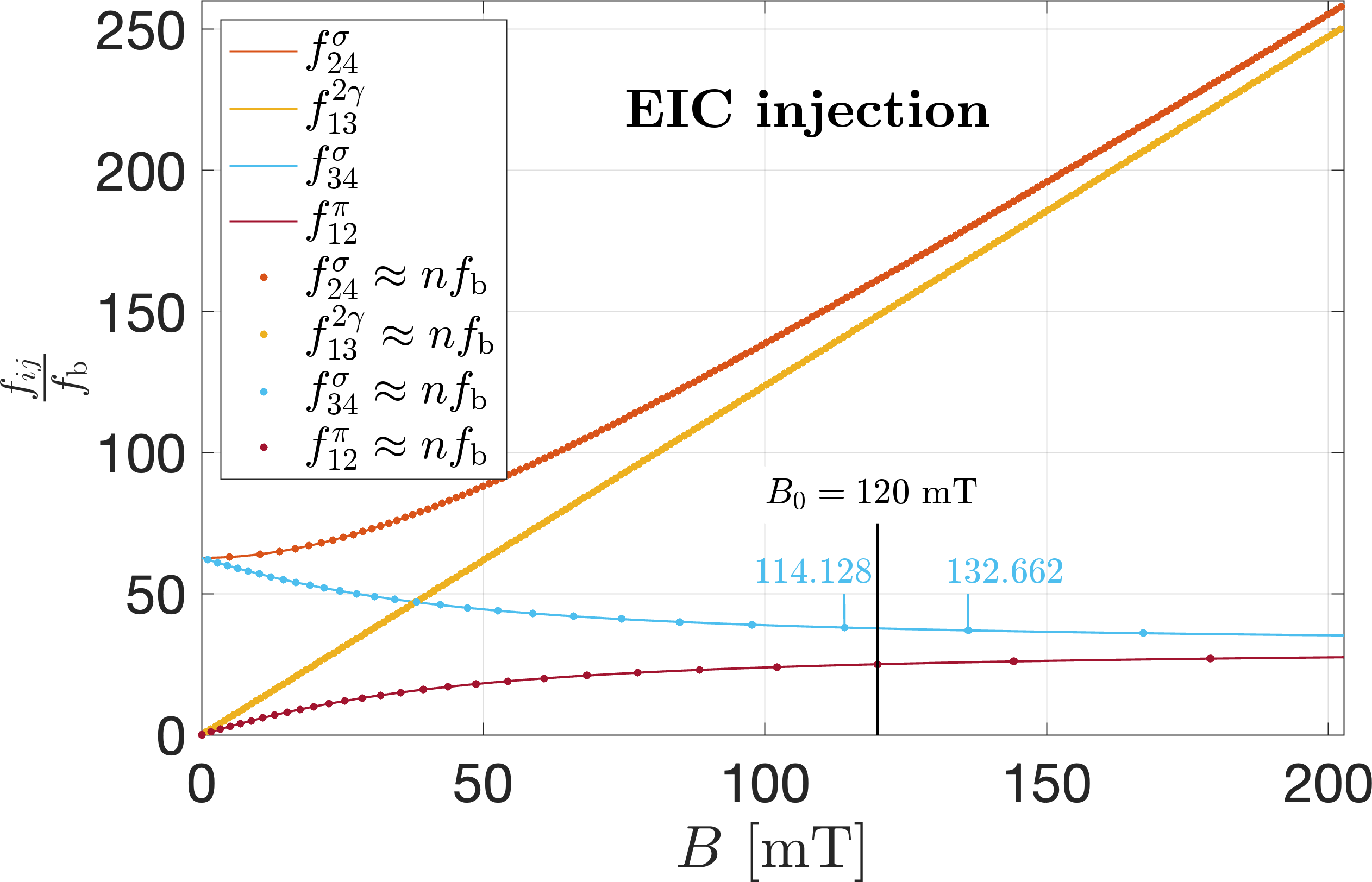}
		\caption{EIC injection (23.5 GeV)}
		\label{fig:eic-HFS-injection}
	\end{subfigure}
	\hfill
	\begin{subfigure}[t]{\columnwidth}
		\centering
		\includegraphics[width=\columnwidth]{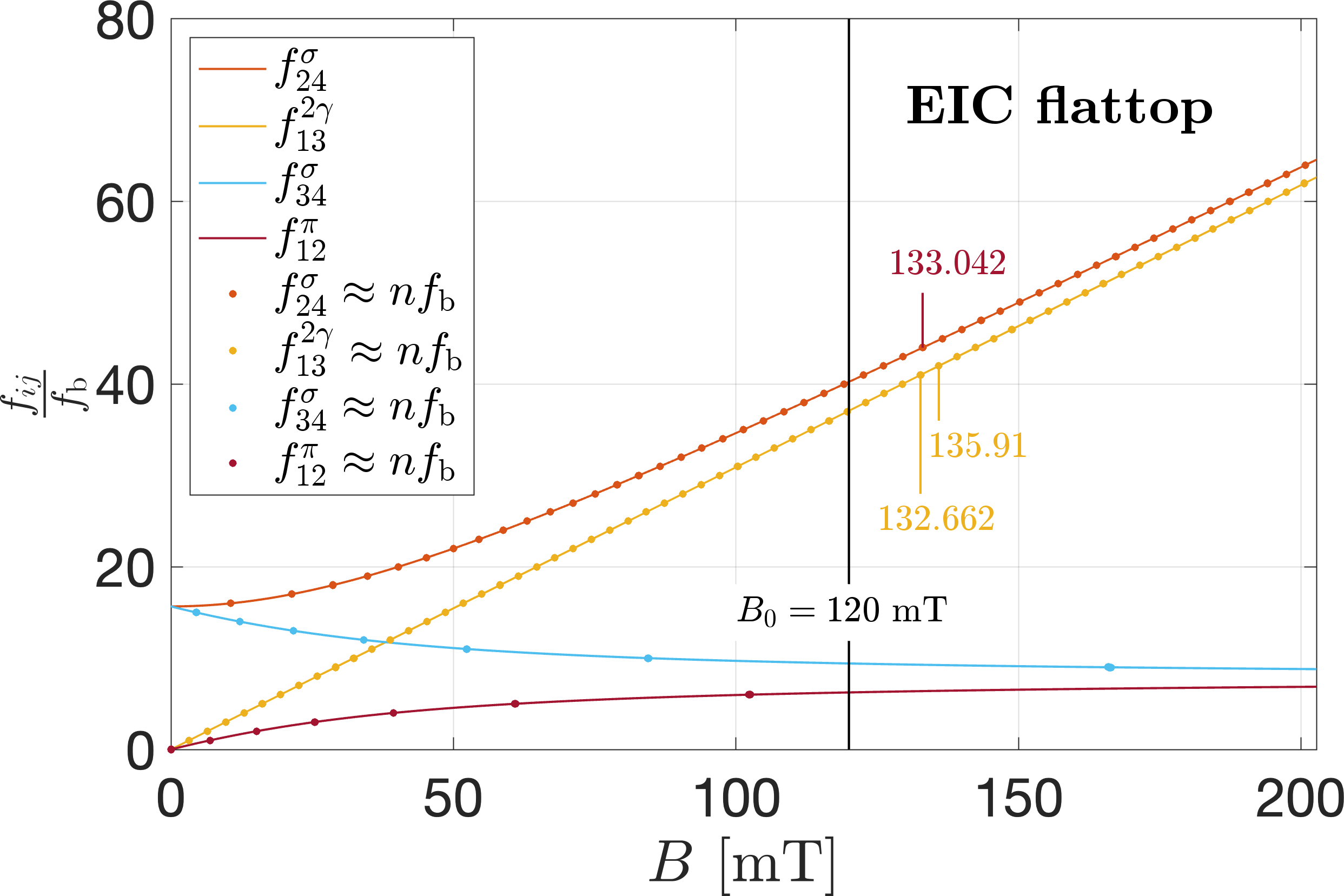}
		\caption{EIC flattop (275 GeV)}
		\label{fig:eic-HFS-flattop}
	\end{subfigure}
	\caption{Resonant overlap between hydrogen hyperfine transition frequencies $f_{ij}(B)$ and the harmonic spectrum of the EIC bunch structure for injection (a) and flattop (b) energies. The plots show the harmonic number $f_{ij}(B)/f_\text{b}$ as a function of magnetic field $B$, with markers indicating points where a near-resonant condition $f_{ij}(B) \approx n f_\text{b}$ is satisfied within a tolerance of 0.002. On flattop, in the region near the static holding field $B_0 = \SI{120}{mT}$ used at RHIC, the spacing between adjacent resonances would be $\approx \SI{1.5}{mT}$.}
	\label{fig:hfs-transitions-harmonics-EIC}
\end{figure*}

\subsection{Hyperfine transition resonances in hydrogen for $B_0 = \SI{120}{mT}$}
\label{sec:eic-resonance}

We now examine how the EIC's higher bunch repetition frequencies affect hyperfine transition resonances  based on the results from the spectral analysis summarized in Table\,\ref{tab:harmonic-cutoff-depol}.

\begin{figure*}[hbtp]
	\centering
	\begin{subfigure}[t]{\columnwidth}
		\centering
		\includegraphics[width=\columnwidth]{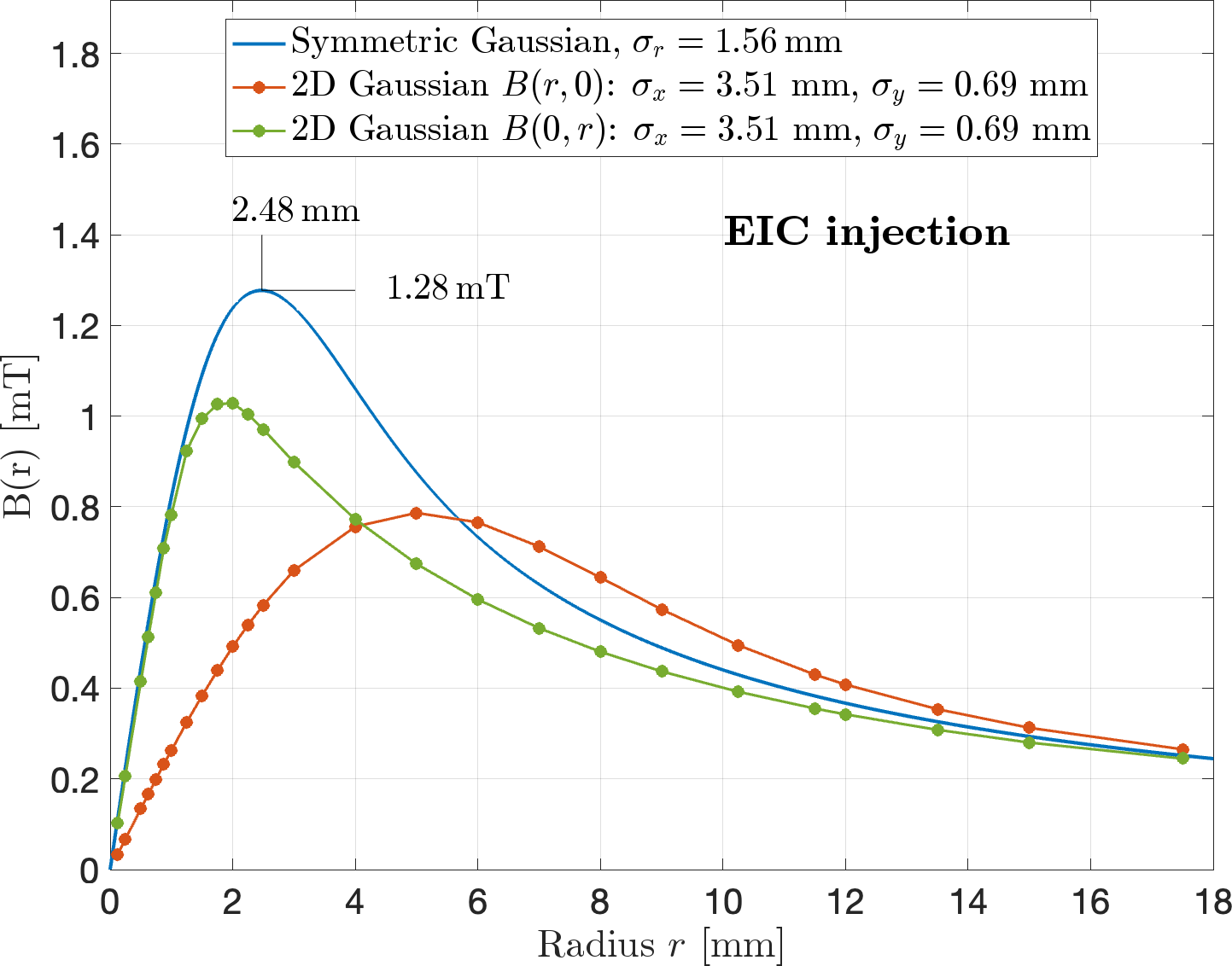}
		\caption{EIC injection (23.5 GeV)}
		\label{fig:EIC-round-vs-asymm-beam-injection}
	\end{subfigure}
	\hfill
	\begin{subfigure}[t]{\columnwidth}
		\centering
		\includegraphics[width=\columnwidth]{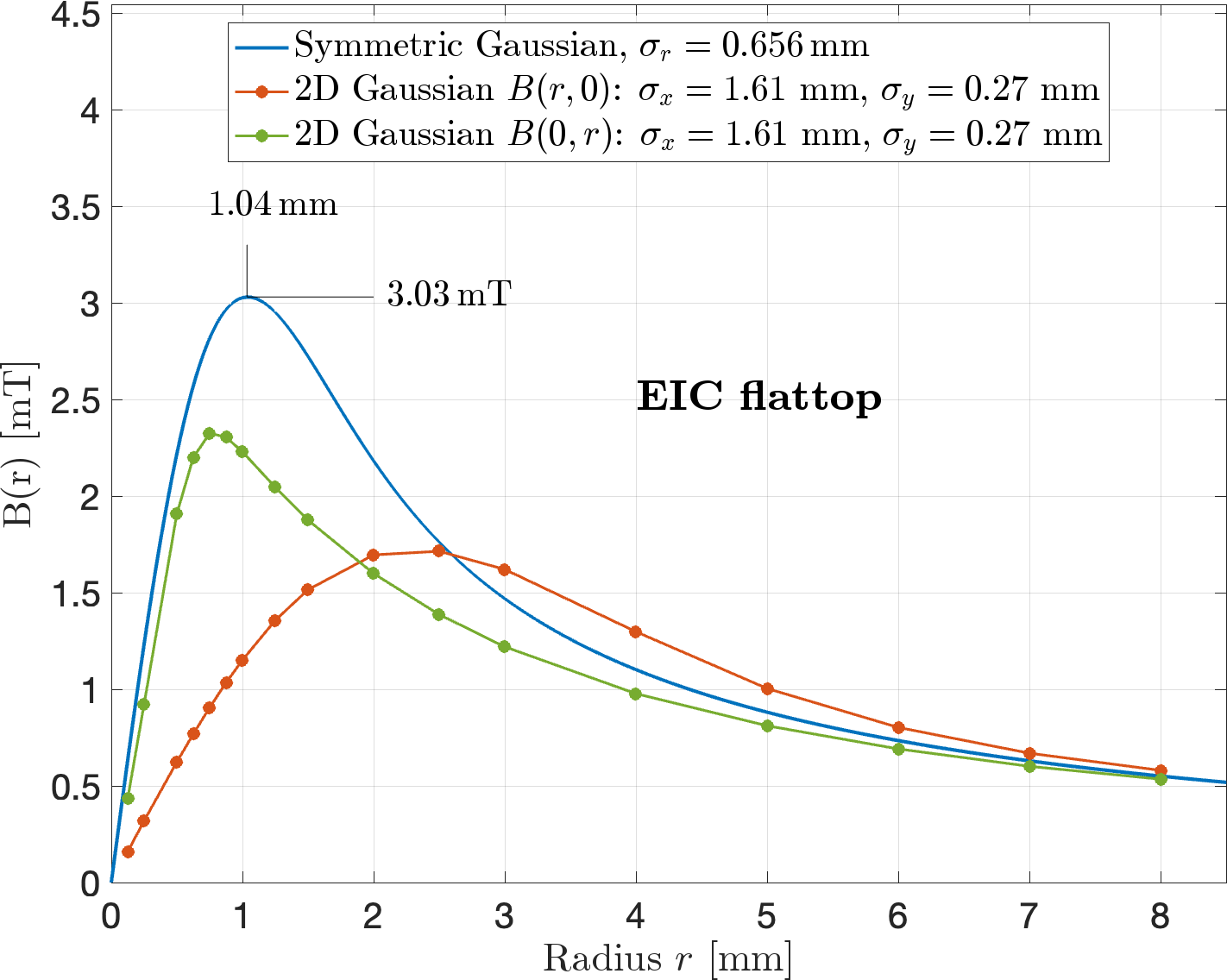}
		\caption{EIC flattop (275 GeV)}
		\label{fig:EIC-round-vs-asymm-beam-flattop}
	\end{subfigure}
	\caption{Magnetic field distribution for the EIC at injection (left panel) and flattop (right). The blue curves show the analytical solutions for a symmetric Gaussian beam with $\sigma_r$ as indicated in the legend, reaching the shown peak fields. Red and green markers show numerical calculations using the Green's function method for an asymmetric beam with $\sigma_x$ and $\sigma_y$ (see legend), along the $x$ and $y$ axes, respectively.}
	\label{fig:EIC-round-vs-asymm-beam}
\end{figure*}

The resonance condition for transitions between hydrogen hyperfine states given in Eq.\,\eqref{eq:harmonic-match} applies at the EIC as well. However, the higher bunch repetition frequencies at the EIC compared to RHIC cause all resonances to shift toward lower harmonic numbers $n = f_{ij}(B)/f_\text{b}$. At RHIC, several transitions -- such as the $\ket{2} \rightarrow \ket{4}$ and $\ket{1} \rightarrow \ket{3}$ transitions -- appeared only at harmonic numbers above $n \approx 375$ and could therefore be safely ignored in the depolarization analysis. At the EIC, these same transitions are mapped to significantly lower harmonic numbers where the spectral power is still high. As shown in Fig.\,\ref{fig:hfs-transitions-harmonics-EIC}, this effect will be most pronounced for EIC flattop energy where all hyperfine transitions fall within the spectral range of potentially depolarizing harmonics, increasing the number of transitions that must be taken into account. 

A second implication is the significantly reduced magnetic field spacing between adjacent resonances. At RHIC, the separation between relevant depolarizing resonance points near the holding field $B_0 = \SI{120}{mT}$ was about \SI{4}{mT} (see Fig.\,\ref{fig:hfs-transitions-harmonics-RHIC}). At the EIC, this spacing compresses to approximately \SI{1.5}{mT} in the same field region [Fig.\,\ref{fig:eic-HFS-flattop}]. This narrow spacing increases the sensitivity of the atomic beam to even modest perturbations of the magnetic field in the vicinity of the interaction region. In particular, beam-induced time-dependent magnetic fields $B_\text{beam}(x,y, t)$ may drive atoms locally and transiently into resonance -- an effect that was negligible at RHIC but must be assessed explicitly for the EIC. The following section addresses this by quantifying the magnitude and spatial variation of beam-induced magnetic fields at the EIC target.

\subsection{Beam-induced magnetic fields at the EIC target location in IP4}
\label{sec:eic-bfield}

In order to relate the magnetic field distribution of an elliptic beam to that of an equivalent round beam, we define first a circular beam profile with the same RMS transverse area. This is achieved by equating the area $\pi \sigma_x \sigma_y$ of the original Gaussian beam with the area of a symmetric beam $\pi \sigma_r^2$, yielding
\begin{equation}
	\sigma_r = \sqrt{\sigma_x \sigma_y}\,.
\end{equation}
This effective round-beam size corresponds to using geometric means of the normalized emittances and beta functions,
\begin{equation}
	\sigma_r = \sqrt{ \frac{\epsilon_\text{avg}^\text{n} \cdot \beta_\text{avg}}{\beta\gamma} }\,,
\end{equation}
with
\begin{equation}
	\epsilon_\text{avg}^\text{n} = \sqrt{\epsilon_x^\text{n} \epsilon_y^\text{n}}, \quad \text{and} \quad
	\beta_\text{avg} = \sqrt{\beta_x \beta_y}\,,
\end{equation}
ensuring that the round-beam approximation preserves both the total charge density and transverse extent relevant for calculating average magnetic fields.

The peak instantaneous magnetic flux densities are calculated using the same methodology as in Section\,\ref{sec:Bfield-spatial-RHIC}. For a round Gaussian beam with $\sigma_x = \sigma_y = \sigma_r$, the magnetic field follows the analytical form previously described in Eq.\,\eqref{eq:B_r_with_F}. For the asymmetric Gaussian beam parameters of the EIC at IP4, listed in Table\,\ref{tab:bunch-params}, we employ the vector potential approach described in Section \ref{sec:elliptic-beam} to numerically calculate the magnetic field.

Figure\,\ref{fig:EIC-round-vs-asymm-beam} compares the magnitude of the magnetic field as a function of distance $r$ from the center of the current distribution for both the symmetric approximation and the full asymmetric calculation. The magnetic field is plotted along the $x$ and $y$ axes, parallel to the long and short axes of the elliptical beam current distribution, respectively. Unlike the round beam case where the field is purely azimuthal with equal magnitude at fixed radius, the asymmetric beam produces different field distributions when measured along these principal axes. Notably, the magnetic field magnitude of the asymmetric current distribution does not exceed that of the equivalent round beam at any radius. This indicates that the round beam approximation provides a safe conservative upper limit for the expected magnetic flux density in the vicinity of the beam.

\subsection{Quantum mechanical depolarization analysis}
\label{subsec:QM-EIC-analysis}

The preceding analysis has shown that EIC operation at $B_0 = \SI{120}{\milli\tesla}$ brings hyperfine transitions into the range of populated beam harmonics, creating potential depolarization risks. As illustrated in Fig.\,\ref{fig:hfs-transitions-harmonics-EIC}, the EIC's higher bunch frequency maps hyperfine transitions to much lower harmonic numbers compared to RHIC. For flattop operation, this creates problematic resonance scenarios: the $\sigma_{24}$ and two-photon $f_{13}^{2\gamma}$ transitions exhibit extremely dense spacing of approximately \SI{1.5}{\milli\tesla}, while power broadening effects can significantly widen effective resonance regions.

To quantify these effects, we use the quantum mechanical framework from Appendix~\ref{app:QM}. The stimulated transition rate for a specific hyperfine transition is
\begin{equation}
	\Gamma_{ij}(f) = \frac{2\pi}{\hbar} |\langle j|H_1|i\rangle|^2 S(f) V_{\text{int}}\,,
\end{equation}
where $S(f) = B_1(f)^2/(2\mu_0)$ is the spectral power density and the matrix elements depend on the transition type through the Breit-Rabi mixing coefficients.

\begin{figure}[tbp]
	\centering
	\includegraphics[width=\columnwidth]{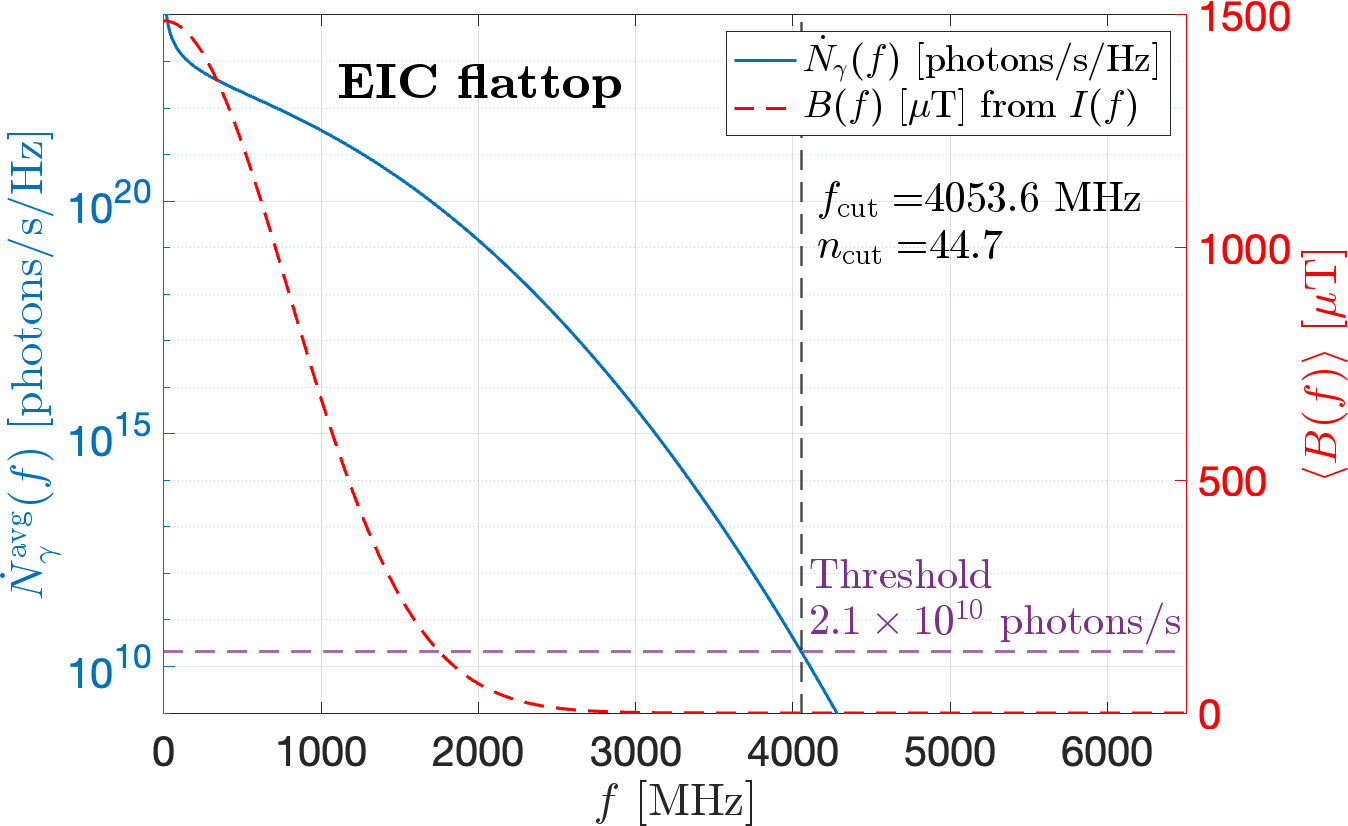}
	\caption{Photon emission rate and magnetic field spectral envelope for EIC flattop operation (275 GeV). The blue solid line shows the average photon emission rate $N_\gamma^{\text{avg}}(f)$ (left axis), with the photon emission threshold at $2.1 \times 10^{10}$ photons/s corresponding to $f_{\text{cut}} = \SI{4053.6}{\mega\hertz}$ and harmonic number $n_{\text{cut}} = 44.7$. The red dashed line shows the beam-induced magnetic field spectral envelope $B(f)$ (right axis), averaged over the interaction region.}
	\label{fig:EIC-photon-Bfield-envelope}
\end{figure}

Consider the $\pi_{12}$ transition resonance at $B_0 = \SI{102}{\milli\tesla}$. From Fig.\,\ref{fig:eic-HFS-flattop}, this occurs at harmonic number $n = 6$ (frequency $f = \SI{544}{\mega\hertz}$). The EIC beam spectral envelope (see Fig.\,\ref{fig:EIC-photon-Bfield-envelope}) provides $B_1 = \SI{1174}{\micro\tesla}$ at this frequency. At $B_0 = \SI{102}{\milli\tesla}$, the dimensionless field parameter $x = 2.01$ gives $\cos^2\theta = 0.946$ for this $\pi$-transition. The Rabi frequency is $\Omega = \SI{2.01e8}{\radian\per\second}$, yielding a transition probability
\begin{equation}
	\Pi = \sin^2\left(\frac{\Omega \tau_{\text{int}}}{2}\right) = \sin^2(1708) \approx 0.73\,.
\end{equation}
This demonstrates that 73\% of hydrogen atoms undergo hyperfine transitions when encountering this resonance. Such a dramatic depolarization effect would be immediately visible in the BRP.

As a second example, consider the $\sigma_{24}$ transition at $B_0 = \SI{119.1}{\milli\tesla}$ and harmonic number $n = 40$ from Fig.\,\ref{fig:eic-HFS-flattop} (frequency $f = \SI{3627}{\mega\hertz}$). For this case, we use the spatial field distribution approach with an effective field amplitude $B_1 = \SI{1.5}{\milli\tesla}$ from Fig.\,\ref{fig:EIC-round-vs-asymm-beam-flattop}. The interaction time is set by the duration atoms spend traversing the localized high-field region, given by $\tau_{\text{int}} = \SI{1.5}{\milli\meter} / \SI{1807}{\meter\per\second} \approx \SI{0.83}{\micro\second}$. The mixed $\sigma$-transition has a matrix element of $0.130$ at this field strength, yielding a Rabi frequency $\Omega = \SI{9.51e7}{\radian\per\second}$ and a transition probability of approximately $36\%$. These calculations demonstrate that EIC operation in the 120 mT region leads to unavoidable depolarization effects. While the extreme case of operating directly on resonance (73\% depolarization) can be avoided through proper $B_0$ field selection, the dense resonance spacing of approximately \SI{1.5}{mT} means that beam-induced field variations will sweep atoms across multiple resonance conditions. Power broadening effects (Sec\,\ref{sec:theoretical-framework-power=broadening}) further widen each resonance by approximately $\pm \SI{0.3}{mT}$, increasing the probability of resonant encounters and causing significant polarization loss (36\% demonstrated here). The quantum mechanical analysis validates the need for alternative operating conditions that move all hyperfine transitions away from populated beam harmonics where such encounters become unavoidable.

These quantum mechanical calculations provide valuable physical insight but represent order-of-magnitude estimates rather than precise predictions. The analysis assumes uniform conditions, whereas the actual beam-induced fields exhibit complex temporal structure and strong spatial variation across the atomic beam volume. The calculated probabilities demonstrate physical capability for significant depolarization rather than quantitative forecasts.

\subsection{From RHIC to EIC: increasing HJET holding field to suppress depolarizing resonances}
\label{sec:EIC-flattop-conclusions}

As discussed in Section\,\ref{sec:eic-resonance}, the use of a static holding field of $B_0 = \SI{120}{mT}$, as employed at RHIC, would be incompatible with reliable operation at the EIC. At this field strength, essentially all hyperfine transitions in hydrogen would lie within the dense spectrum of beam-induced harmonics, leading to significant depolarization.

The critical harmonic cutoff for depolarizing photon emission at EIC flattop lies around $f_{\text{cut}} = \SI{4054}{\mega\hertz}$ (harmonic number $n_{\text{cut}} \approx 45$), as shown in Fig.\,\ref{fig:EIC-photon-Bfield-envelope}. While this cutoff is comparable to RHIC in terms of harmonic number, the EIC's higher bunch frequency ($f_b = \SI{90.683}{\mega\hertz}$) maps hyperfine transitions to much lower harmonic numbers than at RHIC. At $B_0 = \SI{120}{mT}$, virtually all transitions become vulnerable to resonant depolarization, as indicated in Fig.\,\ref{fig:eic-HFS-flattop}.

Exacerbating this issue, the magnetic field generated by the beam itself, on the order of \SI{3}{mT} as shown in Fig.\,\ref{fig:EIC-round-vs-asymm-beam-flattop}, further compromises target operation. Given the narrow resonance spacing of approximately \SI{1.5}{\milli\tesla} under these conditions, such beam-induced field variations can sweep atoms across multiple hyperfine resonances, making target operation at \SI{120}{mT} untenable.

\begin{figure}[hbt]
	\centering
	\includegraphics[width=\columnwidth]{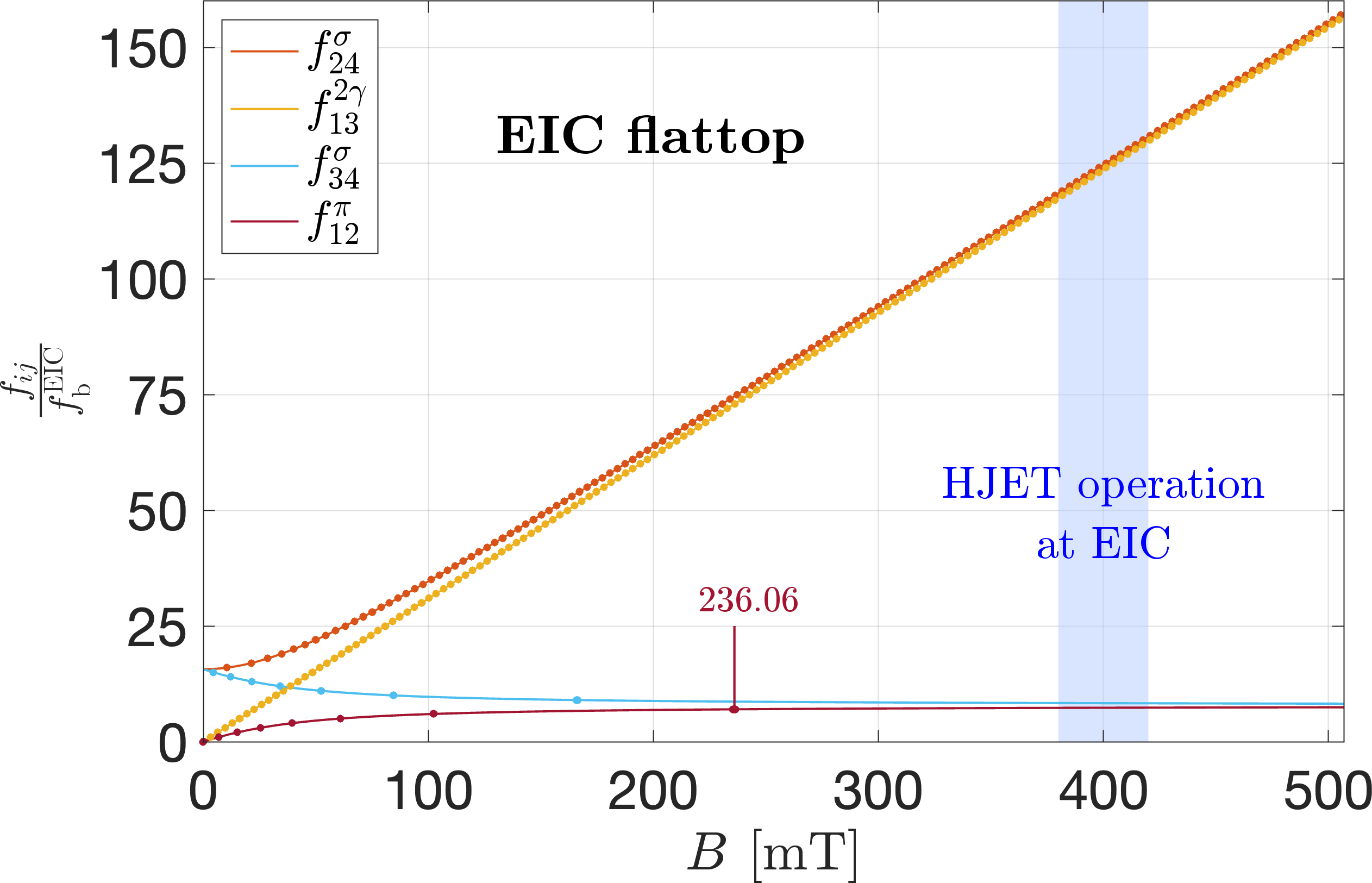}
		\caption{Solution for EIC is to operate HJET in the vicinity in the blue shaded region at a magnetic guide field of $B_0 \approx \SI{400}{mT}$. The highest magnetic field where the rightmost resonance for $f^\pi_{14}$ occurs is indicated.}
		\label{fig:eic-HFS-flattop-solution}
\end{figure}

The solution suggested here is to increase the holding field to eliminate resonance overlap. Figure\,\ref{fig:eic-HFS-flattop-solution} illustrates that above $B_0 \approx \SI{236.06}{mT}$, the highest transition frequencies $f_{12}^\pi$ and $f^\sigma_{34}$ no longer coincide with any harmonic that could induce depolarization, as harmonic number 8 is never reached by either \( f_{12}^\pi \) or \( f^\sigma_{34} \) beyond this field. Operating the HJET in the vicinity of \( B_0 \approx \SI{400}{mT} \), e.g., in the blue shaded region shown in Fig.\,\ref{fig:eic-HFS-flattop-solution}, ensures a region free from depolarizing conditions, and will keep all hyperfine transition frequencies at least a factor of 
\begin{equation}
	\frac{f_{13}}{f_\text{b}} \approx \frac{125}{n_\text{cut}} \approx 2.8
\end{equation}
away from populated beam harmonics. This configuration appears feasible for both EIC injection and flattop energies and provides a reliable solution for suppressing beam-induced depolarization.

The quantum mechanical analysis presented in Section\,\ref{subsec:QM-EIC-analysis} provides additional validation of these concerns, demonstrating that when resonance conditions are encountered at a magnetic guide field of \SI{120}{mT} in the EIC, significant target depolarization occurs (up to 73\% for direct resonance hits, $15-35\%$ for spatial field effects). However, while the quantum mechanical analysis demonstrates the physics underlying these depolarization risks, the primary justification for the \SI{400}{mT} recommendation remains the photon emission threshold analysis, which provides a more robust framework for handling the broadband, spatially varying RF fields characteristic of bunched beam environments.

Beyond eliminating depolarizing resonances, operating at 400 mT provides substantial improvements in systematic uncertainties from beam-induced field asymmetries. To quantify this additional benefit, we analyze the polarization asymmetries using the methodology established for RHIC in Eq.\,\eqref{eq:rel_pol_asymmetry-RHIC} and compare the three operational scenarios.

The beam-induced magnetic field characteristics for the three operational scenarios are summarized in Table\,\ref{tab:beam_fields}, with peak field values extracted from Figs.\,\ref{fig:Bfield-beam-RHIC} (RHIC) and \ref{fig:EIC-round-vs-asymm-beam} (EIC). The field offset values, representing the average magnetic field asymmetry across the atomic beam radius in horizontal direction, are calculated by integrating the azimuthal magnetic field over the respective beam cross-sections using the transverse beam sizes ($\sigma_r$) listed in the table.

The analysis demonstrates that the 400 mT holding field provides a dual benefit: complete elimination of depolarizing resonances while reducing the systematic uncertainties from beam-induced magnetic field  by more than an order of magnitude compared to RHIC operation. Both EIC scenarios exhibit polarization asymmetries well below 0.1\%, representing improvements of $18\times$ (injection) and $8\times$ (flattop) relative to RHIC for this effect, confirming that the higher field strength addresses both operational reliability and precision requirements for EIC polarimetry.

\begin{table}[htbp]
	\centering
	\normalsize
	\caption{Beam-induced magnetic field parameters and resulting polarization asymmetries for RHIC and EIC operational scenarios. The table shows the static holding field $B_0$, the dimensionless field parameter $x = B_0/B_c$, the nuclear polarization for combined injection of states $|1\rangle + |4\rangle $ or $|2\rangle + |3\rangle $,  the transverse beam size $\sigma_r$, and the peak beam-induced field $B_{\text{max}}$. The offset represents the average field asymmetry calculated by integrating over the left and right halves of the beam cross-section. The effective magnetic fields in the left and right hemispheres are given by $B_{\text{L}}$ and $B_{\text{R}}$, respectively, with $\Delta B$ being the total field difference. The target polarization asymmetry $\delta Q/Q$ is calculated using Eq.\,\eqref{eq:rel_pol_asymmetry-RHIC}.}
	\label{tab:beam_fields}
	\begin{ruledtabular}
	\begin{tabular*}{\columnwidth}{@{\extracolsep{\fill}}p{1.5cm}rrrr}
		&  & RHIC at IP\,12& \multicolumn{2}{c}{EIC at IP\,4} \\
		Parameter & Unit & flattop & injection & flattop \\
		\hline
		Energy & GeV & 255 & 23.5 & 275 \\
		$B_0$ & mT & 120 & 400 & 400 \\
		$x$ & -- & 2.4 & 7.9 & 7.9 \\
		$|Q_{|1\rangle+|4\rangle}|$  & \multirow{2}{*}{--} & \multirow{2}{*}{0.962} & \multirow{2}{*}{0.996} & \multirow{2}{*}{0.996} \\ 
$|Q_{|2\rangle+|3\rangle}|$  & & &  &   \\

		$\sigma_r$ & mm & 0.23 & 1.57 & 0.66 \\
		$B_{\text{max}}$ & mT & 2.73 & 1.28 & 3.03 \\
		Offset & mT & 2.09 & 0.98 & 2.32 \\
		$B_{\text{L}}$ & mT & 122.1 & 401.0 & 402.3 \\
		$B_{\text{R}}$ & mT & 117.9 & 399.0 & 397.7 \\
		$\Delta B$ & mT & 4.2 & 2.0 & 4.6 \\
		$\left(\frac{\delta Q}{Q}\right)$ & \% & 0.253 & 0.012 & 0.027 \\
	\end{tabular*}
	\end{ruledtabular}
\end{table}
 
\section{Conclusion and Outlook}
\label{sec:conclusion}

This work has systematically investigated the risk of beam-induced depolarization in the hydrogen jet polarimeter system, with a focus on identifying conditions under which the system can function reliably at the Electron-Ion Collider (EIC). Polarization measurements are essential at both injection and flattop energies, and the goal has been to define operational settings for the magnetic holding field $B_0$ that ensure immunity from depolarizing resonances.

A realistic model of the atomic hyperfine level structure under magnetic fields was combined with a detailed description of the beam’s temporal structure to assess potential depolarization mechanisms. In particular, the beam’s bunch structure was treated as a periodic train, allowing for harmonic decomposition and frequency-domain analysis via discrete Fourier transform. This approach provides a rigorous and transparent framework for identifying resonance conditions between beam harmonics and hyperfine transitions, offering a systematic basis for evaluating depolarization risks in beam-target interactions for the EIC. The approach described here can be readily applied to evaluate the situation of the planned  polarized jet target at the LHC\,\cite{LHCspin:2025lvj}.

A key innovation introduced in this study is the formulation of beam-induced depolarization in terms of a photon emission threshold: a cutoff frequency $f_\text{cut}$, above which the likelihood of resonant transitions is significantly reduced due to the steep falloff in spectral power. This provides a robust basis for comparing different accelerator configurations on the same quantitative footing. For RHIC, this cutoff lies near \SI{441.5}{MHz}, corresponding to a harmonic number $n_\text{cut} \approx 47$. At the EIC, due to its approximately 10 $\times$  higher bunch frequency, the same $n_\text{cut}$ corresponds to an absolute cutoff frequency of \SI{4.05}{GHz}. As a result, the same set of hyperfine transitions is exposed to lower harmonic numbers at the EIC, increasing the likelihood of resonant overlap with populated beam harmonics at a given holding field.

To validate this photon emission framework, a rigorous quantum mechanical analysis using proper Breit-Rabi matrix elements and stimulated transition rates was performed. The quantum mechanical calculations demonstrate that when resonance conditions are encountered at the EIC, significant depolarization occurs ($> 70\%$ for direct resonance encounters), while the same transitions at RHIC fall in spectral regions with negligible field amplitudes. This quantum mechanical validation confirms that the photon emission approach correctly identifies problematic frequency ranges, though the simplified treatment of field coherence and spatial uniformity in this approach means these calculations should be viewed as physics demonstrations rather than precise quantitative predictions.

Furthermore, the spatial variation of the magnetic field near the beam but within the target volume was calculated using the Biot-Savart law applied to a two-dimensional Gaussian beam profile. The derivation employed the magnetic vector potential to accurately capture the azimuthal field generated by elliptic beam distributions. This beam-induced field adds asymmetrically to the static holding field, leading to spatial variations in the net magnetic field direction, which can symmetrically alter the local spin orientation and thus the actual nuclear polarization of atoms across the target volume.

All modeling and analysis techniques were benchmarked using parameters from RHIC at flattop, where successful beam polarimetry using the HJET has been demonstrated. The same methods were then applied to EIC conditions, both at injection and flattop energies. It was shown that the current RHIC operating point at $B_0 = \SI{120}{\milli\tesla}$ is no longer viable at the EIC, as nearly all hyperfine transitions would be exposed to populated harmonics in the beam spectrum. A viable solution is to operate the HJET at the EIC at IP\,4 at a significantly higher magnetic field of $B_0 = \SI{400}{\milli\tesla}$. This field setting ensures a clean separation between transition frequencies and harmonic content, providing a buffer of about a factor of three above the depolarization threshold, and is compatible with EIC operation at both injection and flattop energies, ensuring safe, depolarization-free operation of the HJET.

The developed tools enable predictive estimates of depolarizing conditions and support the selection of holding fields and operating modes for the polarized hydrogen target as an absolute beam polarimeter at the EIC and elsewhere. While the analysis centers on hydrogen, the methodology is directly applicable to deuterium, whose more complex hyperfine structure may lead to different resonance conditions and warrants future investigation. To achieve the stringent 1\% relative polarization uncertainty required by the EIC physics program, several additional developments should be pursued: continuous monitoring of the molecular content in the hydrogen jet (rather than infrequent measurements), and implementation of a magnetic guide field system that enables direct measurement of all polarization components of beam polarization vector $\vec P$. These enhancements, combined with the optimized magnetic holding field identified in this work, will establish a robust foundation for high-precision absolute beam polarimetry at the EIC.

\section*{Acknowledgements}
The authors acknowledge useful discussions with Christoph Montag, Kolya Nikolaev, and Anatoli Zelensky, and thank Andrei Poblaguev for his helpful comments.

\bibliographystyle{apsrev4-2}
\bibliography{polarimetry_eic}	
	
\appendix

\section{Molecular contamination in atomic beams}
\label{app:mol-to-atoms}

Atomic beam sources using sextupole magnets inevitably produce a fraction of molecules that do not originate from the nozzle, but rather from defocused atoms that recombine on the inner surfaces of the sextupole magnets. These recombined molecules form an effusive molecular beam that accompanies the focused atomic beam on its way to the target region\,\cite{NASS2003633, Nass:2002xj}. In order to quantify this effect, we analyze in the following data from the ANKE experiment at COSY\,\cite{MIKIRTYCHYANTS201383} where we explicitly wanted to determine the molecular content in the interaction region and its spatial behavior. The approach used in Ref.\,\cite{POBLAGUEV2020164261} to determine the molecular fraction by simply turning off the dissociator is ill-fated, as it does not produce defocused atoms and as such does not lend itself as a method to realistically estimate the molecular content in the target.

The dissociator design developed for the atomic beam source of the ANKE experiment at COSY\,\cite{MIKIRTYCHYANTS201383} was directly adopted for the polarized atomic beam source used in the RHIC HJET\,\cite{Zelenski2005}. The construction drawings were provided by the J\"ulich group and the dissociator design is identical in both systems. As reported in\,\cite{MIKIRTYCHYANTS201383}, degree of dissociation measurements were carried out with a quadrupole mass spectrometer movable on an $xy$ table that allowed determination of the spatial dependence of the molecular to atomic content in the beam some distance (\SI{567}{mm} and \SI{697}{mm}) behind the exit of the last sextupole magnet. The analysis presented here examines the degree of dissociation data obtained, shown in panels (c) and (d) of Fig.\,24 of Ref.\,\cite{MIKIRTYCHYANTS201383}.

\begin{figure}[htb]
	\centering
	\begin{subfigure}[b]{\columnwidth}
		\includegraphics[width=\textwidth]{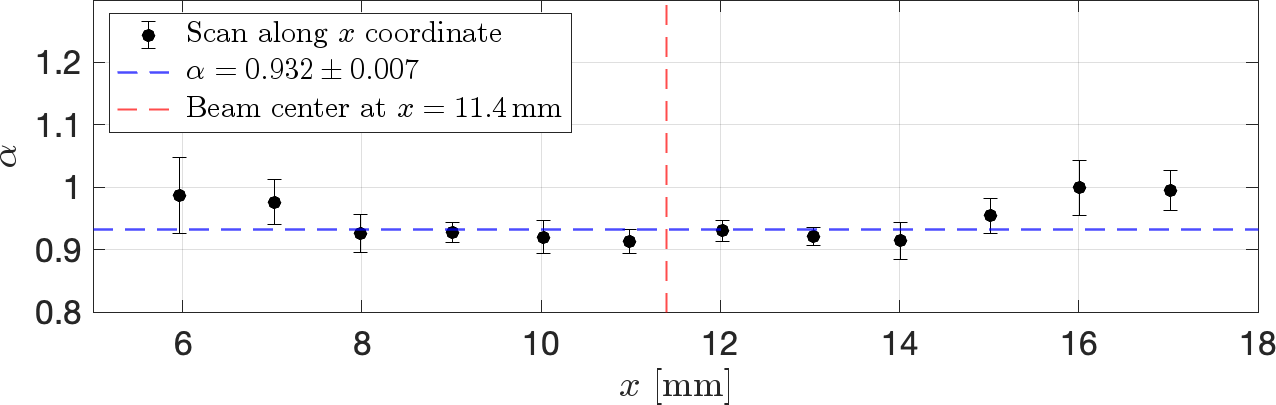}
		\caption{Scan along $x$}
		\label{fig:alpha_x}
	\end{subfigure}
\vspace{0.2cm}
	\begin{subfigure}[b]{\columnwidth}
		\includegraphics[width=\textwidth]{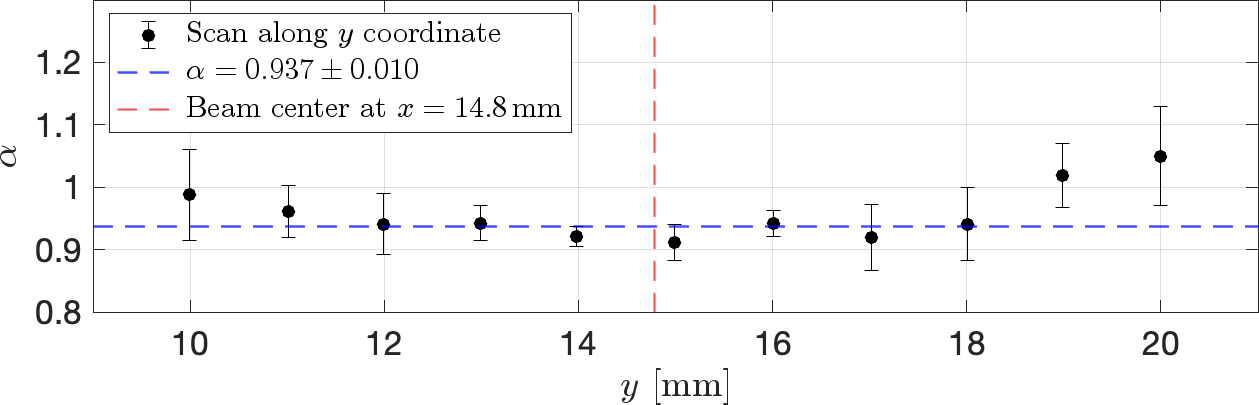}
		\caption{Scan along $y$}
		\label{fig:alpha_y}
	\end{subfigure}
	\caption{Degree of dissociation $\alpha$ measured across orthogonal spatial profiles of the atomic beam. The red dashed lines indicate the beam centers, while the blue dashed lines show constant fits to the data. Subfigure (a) shows the $x$-profile with beam center at $x = 11.4\,\mathrm{mm}$ and fitted constant $\alpha_x = 0.932 \pm 0.007$. Subfigure (b) shows the $y$-profile with beam center at $y = 14.8\,\mathrm{mm}$ and fitted constant $\alpha_y = 0.937 \pm 0.010$.}
	\label{fig:dissociation_profiles}
\end{figure}

The degree of dissociation $\alpha$ was measured at multiple positions along the transverse $x$ and $y$ directions perpendicular to the atomic beam 697 mm behind the exit of the last sextupole magnet. Figure~\ref{fig:dissociation_profiles} shows the results of these measurements along with constant fits to the data. The results demonstrate a flat dependence of $\alpha$ near the beam center, and we confine our analysis to data within $\pm 5$ mm around the beam center since the atomic beam of the HJET at RHIC has a diameter of approximately \SI{10}{mm}\,\cite{Zelenski2005}. For the $x$-profile, centered around $x = 11.4\,\mathrm{mm}$, a fitted constant value of $\alpha_x = 0.932 \pm 0.007$ is obtained, and for the $y$-profile, centered at $y = 14.8\,\mathrm{mm}$, $\alpha_y = 0.937 \pm 0.010$. The combined result, calculated as an inverse-variance weighted average of both spatial profiles, gives 
\begin{equation}
	\alpha = 0.934 \pm 0.006\,.
	\label{eq:averaged-alpha}
\end{equation}

The degree of dissociation of the atomic beam is defined in terms of the atomic density $\rho_\text{atom}$ and molecular density $\rho_\text{mol}$ as
\begin{equation}
	\alpha = \frac{\rho_\text{atom}}{\rho_\text{atom} + 2\rho_\text{mol}}\,,
\end{equation}
and this can be rearranged to obtain the molecular-to-atomic density ratio
\begin{equation}
	\frac{\rho_\text{mol}}{\rho_\text{atom}} = \frac{1 - \alpha}{2\alpha}\,.
\end{equation}
Using the measured value of $\alpha$ from Eq.\,\eqref{eq:averaged-alpha}, we obtain
\begin{equation}
 	\frac{\rho_\text{mol}}{\rho_\text{atom}} = 0.035 \pm 0.003\,.
\end{equation}

This result indicates that approximately 3 to 4\% of the target density consists of hydrogen molecules, consistent with findings from studies on similar atomic beam sources\,\cite{NASS2003633, Nass:2002xj}. These unpolarized molecules systematically reduce the target polarization of the HJET.

\section{Hyperfine interaction Hamiltonian and nuclear polarizations for ground state hydrogen}
\label{app:Breit-Rabi-derivation}

The complete Hamiltonian for the ground state hydrogen atom in an external magnetic field $\vec{B} = B \vec{e}_z$, where $\vec{e}_z$ defines the quantization axis, consists of three terms
\begin{equation}
	H = A_{\text{hfs}} \boldsymbol{I} \cdot \boldsymbol{J} - \boldsymbol{\mu}_J \cdot \vec{B} - \boldsymbol{\mu}_I \cdot \vec{B}\,.
\end{equation}
Here $\boldsymbol{I}$ is the nuclear spin operator ($I = \frac{1}{2}$ for hydrogen), $\boldsymbol{J}$ is the total electron angular momentum operator ($J = \frac{1}{2}$ for the ground state), and $A_{\text{hfs}}$ is the hyperfine coupling constant. For the hydrogen ground state (1s), the orbital angular momentum is zero ($l = 0$), so the total electron angular momentum equals the electron spin: $\boldsymbol{J} = \boldsymbol{S}$. The magnetic moment operators are
\begin{equation}
	\boldsymbol{\mu}_J = -g_J \mu_B \boldsymbol{J} \quad \text{and} \quad \boldsymbol{\mu}_I = g_I \mu_N \boldsymbol{I}\,,
\end{equation}
so that when we choose the quantization axis along $\vec{B}$, the complete Hamiltonian becomes
\begin{equation}
	H = A_{\text{hfs}} \boldsymbol{I} \cdot \boldsymbol{J} + g_J \mu_B J_z B + g_I \mu_N I_z B\,.
\end{equation}
The hyperfine coupling constant is related to the zero-field hyperfine splitting by $A_{\text{hfs}} = 4E_{\text{hfs}}/\hbar^2$, where $E_{\text{hfs}}$ is given in Eq.~(8).

We work in the uncoupled basis $\{|m_J, m_I\rangle\}$ where $m_I, m_J = \pm \frac{1}{2}$. The four basis states are labeled in decreasing order of hyperfine energies, as given in Eqs.\,\eqref{eq:HFS-state-desription-1-to-4}. Since the total angular momentum projection $m_F = m_J + m_I$ is conserved by the hyperfine interaction, states with the same $m_F$ can couple while states with different $m_F$ cannot. Therefore, $|1\rangle$ and $|3\rangle$ remain uncoupled, while $|2\rangle$ and $|4\rangle$ (both with $m_F = 0$) form a coupled 2×2 system.

The dot product $\boldsymbol{I} \cdot \boldsymbol{J} = I_z J_z + \frac{1}{2}(I_+ J_- + I_- J_+)$ has diagonal matrix elements $\langle m_J, m_I | I_z J_z | m_J, m_I \rangle = \hbar^2 m_I m_J$ and off-diagonal elements $\langle 2 | I_- J_+ + I_+ J_- | 4 \rangle = \langle 4 | I_- J_+ + I_+ J_- | 2 \rangle = \hbar^2$. The hyperfine matrix is
\begin{equation}
	\frac{\boldsymbol{I} \cdot \boldsymbol{J}}{\hbar^2/4} = \begin{pmatrix}
		1 & 0 & 0 & 0 \\
		0 & -1 & 0 & 2 \\
		0 & 0 & 1 & 0 \\
		0 & 2 & 0 & -1
	\end{pmatrix}
\end{equation}

Using the dimensionless field strength parameter $x$ defined in Eq.\,(6) and defining $y = \frac{2 g_I \mu_N B}{E_{\text{hfs}}}$, we note that $y \ll x$ since $\mu_N/\mu_B = 1/1836.15$ and $g_I/g_J \approx 2.8$. The complete dimensionless Hamiltonian becomes
\begin{widetext}
\begin{equation}
	\frac{H}{E_{\text{hfs}}/4} = \begin{pmatrix}
		1 + 2x + y & 0 & 0 & 0 \\
		0 & -1 + 2x - y & 0 & 2 \\
		0 & 0 & 1 - 2x - y & 0 \\
		0 & 2 & 0 & -1 - 2x + y
	\end{pmatrix}
\end{equation}
\end{widetext}

The eigenvalues can be found by diagonalizing this matrix. States $|1\rangle$ and $|3\rangle$ remain uncoupled with eigenvalues $E_{|1\rangle} = \frac{E_{\text{hfs}}}{4}(1 + 2x + y)$ and $E_{|3\rangle} = \frac{E_{\text{hfs}}}{4}(1 - 2x - y)$. States $|2\rangle$ and $|4\rangle$ couple through the hyperfine interaction with eigenvalues $E_{|2\rangle,|4\rangle} = \frac{E_{\text{hfs}}}{4}\left[y - 1 \pm 2\sqrt{1 + 2xy + x^2}\right]$. For typical magnetic fields where $x \gg y$, this reduces to the familiar Breit-Rabi formula $E_{|2\rangle,|4\rangle} \approx \frac{E_{\text{hfs}}}{4}\left[-1 \pm 2\sqrt{1 + x^2}\right] + \frac{E_{\text{hfs}}}{4}y$. Combining all four eigenvalues and including the nuclear Zeeman correction, we obtain the complete Breit-Rabi energy formula given in Eq.\,\eqref{eq:hyperfine-energies} in the main text.

The nuclear target polarization of each hyperfine state is determined by the quantum mechanical expectation value of the nuclear spin component along the quantization axis, expressed via
\begin{equation}
	Q_{|i\rangle} = \frac{2}{\hbar} \langle \psi_i | I_z | \psi_i \rangle\,.
\end{equation}

To calculate this, we need the explicit eigenvectors. States $|1\rangle$ and $|3\rangle$ remain pure uncoupled states at all field strengths, while states $|2\rangle$ and $|4\rangle$ become mixed states. The corresponding wave functions are
\begin{equation}
	\begin{split}
		|\psi_1\rangle &= |e^\uparrow p^\uparrow\rangle\,, \\
		|\psi_2\rangle &= \cos\theta |e^\uparrow p^\downarrow\rangle + \sin\theta |e^\downarrow p^\uparrow\rangle\,, \\
		|\psi_3\rangle &= |e^\downarrow p^\downarrow\rangle\,, \\
		|\psi_4\rangle &= \cos\theta |e^\downarrow p^\uparrow\rangle - \sin\theta |e^\uparrow p^\downarrow\rangle\,,
	\end{split}
	\label{eq:hfs-states-theta}
\end{equation}
where the mixing angle $\theta$ satisfies $\tan (2\theta) = 1/x$. 
Using the matrix elements $\left\langle \pm \frac{1}{2}, m_J \left| I_z \right| \pm \frac{1}{2}, m_J \right\rangle = \pm \frac{\hbar}{2}$, the nuclear target polarizations for the different states are obtained, and given in Eq.~\eqref{eq:Q-of-x-for1to4} in the main text.

In the weak field limit ($x \to 0$), states $|2\rangle$ and $|4\rangle$ have zero nuclear polarization, reflecting equal superposition of parallel and antiparallel nuclear-electron spin configurations. In the strong field limit ($x \to \infty$), all states approach maximum nuclear polarization ($\pm 1$). 

\section{Quantum mechanical analysis of hyperfine transitions}
\label{app:QM}

This appendix presents the quantum mechanical framework for analyzing beam-induced hyperfine transitions in hydrogen atoms using proper Breit-Rabi matrix elements and Fermi's Golden Rule\,\cite{ramsey1956,CohenTannoudjiQM-German}. While a full time-dependent solution of the hyperfine Hamiltonian would be required to compute exact state populations, the use of Fermi’s Golden Rule is sufficient for identifying resonance conditions and estimating relative transition strengths relevant for depolarization risk assessment.

For hydrogen hyperfine transitions in a magnetic field, the interaction Hamiltonian with the beam-induced RF field follows from Appendix\,\ref{app:Breit-Rabi-derivation}, where the electronic coupling dominates and we have $H_1 \approx -g_J \mu_B \boldsymbol{J} \cdot \vec{B}_1(t)$. The Breit-Rabi eigenstates at field $B_0$ are given in Eq.\,\eqref{eq:hfs-states-theta}, where states $|2\rangle$ and $|4\rangle$ become mixed superpositions of uncoupled spin configurations as derived in Appendix\,\ref{app:Breit-Rabi-derivation}, and the mixing angle satisfies $\tan(2\theta) = 1/x$ with the dimensionless field strength parameter $x$ from Eq.\,\eqref{eq:def-of-x}.

The transition matrix elements depend on the orientation of the beam-induced RF field $\vec{B}_1$ relative to the static holding field $\vec{B}_0$. As detailed in Section\,\ref{subsec:transition-frequencies}, there are six allowed single-photon transitions between the four hyperfine states, classified according to the RF field orientation and selection rules: $\pi$-transitions ($\vec{B}_1 \perp \vec{B}_0$) with $\Delta F = 0, \Delta m_F = \pm 1$, and $\sigma$-transitions ($\vec{B}_1 \parallel \vec{B}_0$) with $\Delta F = \pm 1, \Delta m_F = 0, \pm 1$. Two-photon transitions ($\Delta m_F = 2$) are forbidden for single-photon processes and require much higher field intensities to become significant.

Using Fermi's Golden Rule, the stimulated transition rate between hyperfine states $|i\rangle$ and $|j\rangle$ is given by
\begin{equation}
	\Gamma_{ij}(f) = \frac{2\pi}{\hbar} |\langle j|H_1|i\rangle|^2 S(f) V_{\text{int}}\,,
	\label{eq:Gamma_ij_appendix}
\end{equation}
where $S(f) = B_1(f)^2/(2\mu_0)$ is the spectral power density and the matrix elements depend on the specific transition and magnetic field strength through the Breit-Rabi mixing coefficients.

The stimulated transition rates between all four hyperfine states form a complete $4 \times 4$ matrix with elements
\begin{equation}
	\Gamma_{ij} = \frac{2\pi}{\hbar} S(f) V_{\text{int}} \begin{pmatrix}
		0 & \cos^2\theta & \frac{1}{4} & \frac{1}{4} \\
		\cos^2\theta & 0 & \cos^2\theta & \frac{x^2}{(1+x^2)^2} \\
		\frac{1}{4} & \cos^2\theta & 0 & \cos^2\theta \\
		\frac{1}{4} & \frac{x^2}{(1+x^2)^2} & \cos^2\theta & 0\,,
	\end{pmatrix}
	\label{eq:transition-matrix-elements}
\end{equation}
where the rows and columns correspond to states $|1\rangle$, $|2\rangle$, $|3\rangle$, and $|4\rangle$, respectively. The diagonal elements are zero since no state can transition to itself under single-photon processes. The off-diagonal elements represent squared matrix elements for different transition types: $\cos^2\theta$ for $\pi$-transitions ($\vec{B}_1 \perp \vec{B}_0$) involving operators $J_\pm = J_x \pm iJ_y$; $1/4$ for pure $\sigma$-transitions ($\vec{B}_1 \parallel \vec{B}_0$) using $J_z$; and $x^2/(1+x^2)^2$ for the mixed $\sigma$-transition $|2\rangle \leftrightarrow |4\rangle$. The matrix includes both single-photon and two-photon transition elements; while the $1\rightarrow3$ and $3\rightarrow1$ transitions are forbidden as single-photon processes ($\Delta m_F = 2$), their matrix elements represent two-photon coupling strengths that are negligible under realistic photon densities. This matrix demonstrates that all hyperfine states are coupled through field-dependent transition rates, making simple two-level approximations inadequate.

The fractional change in nuclear polarization during the atomic transit time $\tau_{\text{int}}$ is given by $\Delta Q/Q = \Gamma_{\text{depol}} \tau_{\text{int}}$, where $\Gamma_{\text{depol}}$ represents the effective depolarization rate from all relevant transitions. This framework enables quantitative assessment of beam-induced depolarization effects under specific operational conditions.

\end{document}